\documentclass[conference]{IEEEtran}

\IEEEoverridecommandlockouts
\usepackage{cite}
\usepackage{amsmath,amssymb,amsfonts}
\usepackage{algorithmic}
\usepackage{graphicx}
\usepackage{textcomp}
\usepackage{xcolor}
\usepackage{multirow}
\usepackage{booktabs}
\usepackage{adjustbox}
\usepackage{amsmath}
\usepackage{pifont}
\usepackage{color}

\newcommand{\xmark}{\text{\ding{55}}}

\def\BibTeX{{\rm B\kern-.05em{\sc i\kern-.025em b}\kern-.08em
    T\kern-.1667em\lower.7ex\hbox{E}\kern-.125emX}}
\begin{document}
\title{Contrastive Enhanced Slide Filter Mixer for Sequential Recommendation
}


\DeclareRobustCommand*{\IEEEauthorrefmark}[1]{%
  \raisebox{0pt}[0pt][0pt]{\textsuperscript{\footnotesize #1}}%
}
\author{
    \IEEEauthorblockN{
        Xinyu Du\IEEEauthorrefmark{1}\textsuperscript{$\dagger$}, 
        Huanhuan Yuan\IEEEauthorrefmark{1}\textsuperscript{$\dagger$}, 
        Pengpeng Zhao\IEEEauthorrefmark{1}\textsuperscript{*}, 
        Junhua Fang\IEEEauthorrefmark{1},
        Guanfeng Liu\IEEEauthorrefmark{2}, 
        Yanchi Liu\IEEEauthorrefmark{3}, 
        \\
        Victor S. Sheng\IEEEauthorrefmark{4}, 
        and
        Xiaofang Zhou\IEEEauthorrefmark{5} 
    }
    \IEEEauthorblockA{
        \IEEEauthorrefmark{1}School of Computer Science and Technology, Soochow University, Suzhou, China\\
        \IEEEauthorrefmark{2}Department of Computing, Macquarie University, Australia \\
        \IEEEauthorrefmark{3}Rutgers University, USA
        \IEEEauthorrefmark{4}Texas Tech University, USA \\
        \IEEEauthorrefmark{5}The Hong Kong University of Science and Technology, China\\
        \{xydu, hhyuan\}@stu.suda.edu.cn
        \{ppzhao, jhfang\}@suda.edu.cn \\
        guanfeng.liu@mq.edu.au
        yanchi.liu@rutgers.edu 
        victor.sheng@ttu.edu
        zxf@cse.ust.hk
    }
}
\maketitle

\begingroup
\renewcommand\thefootnote{$\dagger$}
\footnotetext{They are co-first authors with equal contributions.}
\renewcommand\thefootnote{*}
\footnotetext{Corresponding author}

\endgroup

\begin{abstract}
Sequential recommendation (SR) aims to model user preferences by capturing behavior patterns from their item historical interaction data.
Most existing methods model user preference in the time domain, omitting the fact that users' behaviors are also influenced by various frequency patterns that are difficult to separate in the entangled chronological items. However, few attempts have been made to train SR in the frequency domain, and it is still unclear how to use the frequency components to learn an appropriate representation for the user. 
To solve this problem, we shift the viewpoint to the frequency domain and propose a novel Contrastive Enhanced \textbf{SLI}de Filter \textbf{M}ixEr for Sequential \textbf{Rec}ommendation, named \textbf{SLIME4Rec}. 
Specifically, we design a frequency ramp structure to allow the learnable filter slide on the frequency spectrums across different layers to capture different frequency patterns. 
Moreover, a Dynamic Frequency Selection (DFS) and a Static Frequency Split (SFS) module are proposed to replace the self-attention module for effectively extracting frequency information in two ways. 
DFS is used to select helpful frequency components dynamically, and SFS is combined with the dynamic frequency selection module to provide a more fine-grained frequency division.
Finally, contrastive learning is utilized to improve the quality of user embedding learned from the frequency domain.
Extensive experiments conducted on five widely used benchmark datasets demonstrate our proposed model performs significantly better than the state-of-the-art approaches. Our code is available at https://github.com/sudaada/SLIME4Rec. 
\end{abstract}

\begin{figure}[!t]
	\centering
	{
	\includegraphics[width=1\linewidth]{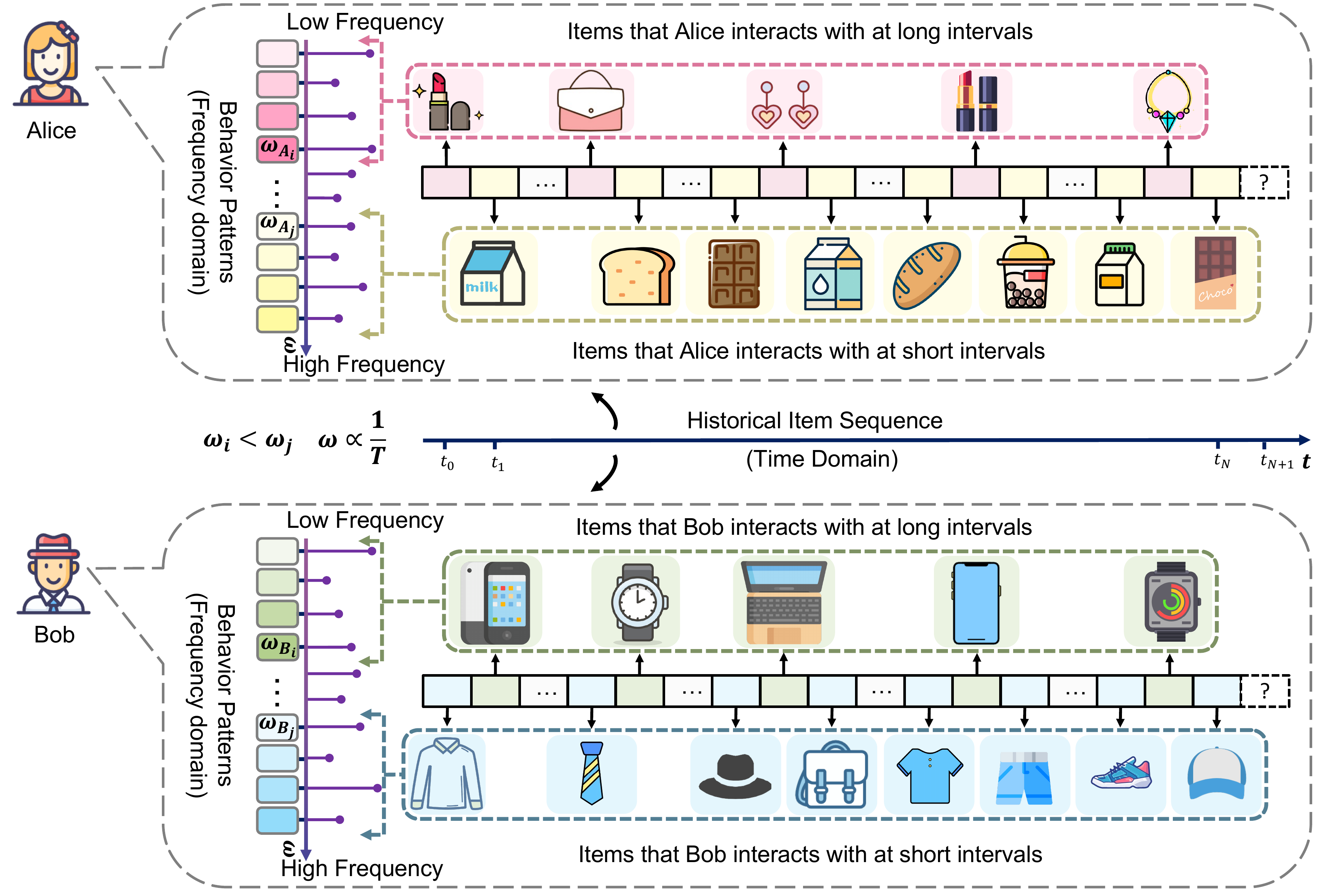}}
	\caption{Illustration of a user's historical item sequence observed from two different perspectives.
	When we observe from the viewpoint of the time domain, the items interacted by Alice and Bob are chronologically ordered along the $t$-axis.
	When we shift our perspective to the frequency domain, the historical item sequence of each user decomposed into multiple behavioral patterns with different frequencies and periods along the $\omega$-axis.
	Both Alice and Bob have low-frequency behavior ($i.e.$, $\omega_{A_{i}},\omega_{B_{i}}$) and high-frequency behavior ($i.e.$, $\omega_{A_{j}},\omega_{B_{j}}$).
	}
	\label{fig_user_item}
\end{figure}

\section{Introduction}
Recommender systems (RSs) have become popular on various online platforms for their ability to accurately recommend a series of proper items that users may be interested in. 
Different from conventional RSs~\cite{BPR_MF} that usually assume user preferences are static, sequential recommendation algorithms consider time series information~\cite{GRU4Rec, Caser, SASRec} to better recognize user behavior patterns from user interaction sequences and capture the evolution of user interests.

The majority of sequential models, however, primarily process historical interactions from the perspective of the time domain and rarely fully exploit the features of users in the frequency domain, which is frequently used in the field of digital signal processing to filter noises~\cite{FMLP-Rec, GFNet}. 
For example, Figure~\ref{fig_user_item} shows the sequence of historical items interacted by Alice and Bob, respectively.
When we observe from the viewpoint of the time domain, all the items are chronologically ordered and intertwined along the $t$-axis, which are in essence noisy or even contain malicious fakes~\cite{FMLP-Rec}. 
By converting the input time features to the frequency domain with a discrete Fourier transform~\cite{DFT1, DFT2}, the historical item sequence of each user can be decomposed into different frequency spectrums to find more fine-grained sequential patterns in different frequencies.

In the time domain, deep recommender models tend to overfit on noisy interaction data~\cite{FMLP-Rec} in the time domain, due to present approaches providing weights to all interacted items in order to uncover user behavior patterns at the item level.
Furthermore, some studies have shown that the widely used self-attention operation is substantially better at gathering global low-frequency information but reduces high-frequency signals~\cite{howVITwork, VIT_seelike_CNN}, which can be easily obtained in the frequency domain.
For this reason, FMLP-Rec~\cite{FMLP-Rec} tries to process sequence data in the frequency domain to attenuate noise information hidden in the feature and capture the frequency patterns for users.
For the frequency-based method, how to effectively employ the frequency components to represent the user's behaviors is a vital problem.
On the one hand, although the model can best preserve the user's thorough information in the frequency domain by simply multiplying a filter that covers all frequency components, it may lead to inferior representations because of noisy inputs~\cite{FEDformer}.
On the other hand, it is not advisable to learn different frequency characteristics simultaneously.
The frequency domain features can be further divided into other components (such as high- and low-frequencies), which could help learn variable patterns for different frequencies. 
Such merit is particularly appealing for the recommendation task, where the user's behaviors tend to show certain periodic trends.
For RSs, the high-frequencies mainly include the items that regularly are brought in a short time interval (e.g., clothes), and the relatively low-frequencies often include the items that usually are purchased in a long time interval (e.g., mobile phone or computer).
In Figure~\ref{fig_user_item}, the behavior $\omega_{B_{j}}$ (e.g., Clothing and Outdoors) is obviously more frequent than $\omega_{B_{i}}$ (e.g., Electronic products) since $\omega_{B_{j}} > \omega_{B_{i}}$. 
But it is hard to recognize these variable patterns when items with all frequencies are entangled in the time sequences.
Hence, flexibly learning high-frequency behaviors (Clothing and Outdoors) and low-frequency behaviors (Electronic products) respectively could be a way to have a better understanding of the whole user preference.
For these reasons, we propose a simple and efficient model named Contrastive Enhanced \textbf{Sli}de Filter \textbf{M}ix\textbf{e}r for Sequential \textbf{Rec}ommendation, or, \textbf{SLIME4Rec} for short. We keep exploring the frequency features in the frequency domain and find that the user historical sequence is composed of a variety of behavior patterns with different periods and frequencies. 
The key component in SLIME4Rec is the filter mixer, which replaces the self-attention module with dynamic frequency selection and static frequency split module in the frequency domain. 
Firstly, different from FMLP-Rec~\cite{FMLP-Rec} which utilizes a global learnable filter to cover all frequency components, a dynamic frequency selection module is proposed to make the learnable filter slide on the frequency domain. 
Specifically, to trade off the different frequency patterns of users, the frequency ramp structure module hierarchically covers a specific frequency range with different neural network layers in the frequency domain.
The main function of this module is to enable SLIME4Rec to adaptively emphasize or filter some frequency components, and ease the overfit phenomenon caused by noise.
Secondly, to ensure all frequency components are considered to best preserve the user's meaningful periodic characteristics, we mix a static frequency split module with the dynamic frequency selection module. It averagely divides all frequency components according to the number of layers, which is to provide a more fine-grained frequency division and recapture frequencies missed by the dynamic frequency selection module.
Finally, to improve the quality of user embedding, we design a contrastive learning task to augment the supervisory signal and take it as a supplement to the recommended task.
The main contributions of this paper are summarized as follows:
\begin{itemize}
    \item We shift the perspective to the frequency domain and design a frequency ramp structure, which uses slide filters to capture different subsets of whole frequency features in different layers. 
    
    \item We propose dynamic frequency selection and static frequency split module, respectively, which mixes different fine-grained frequency features captured by filters of different sizes. 
    
    \item We introduce our frequency feature-based model into the contrastive learning paradigm. By unifying the recommendation task and the user behavior-level self-supervised task, the recommendation performance can be significantly improved.
    
    \item We conduct extensive experiments on five public datasets and demonstrate the superiority of the SLIME4Rec compared to state-of-the-art baselines with less computational complexity.
\end{itemize}

\section{PRELIMINARIES}
Before elaborating on the proposed SLIME4Rec, we first present some necessary notations to formulate the sequential recommendation problem and then introduce some background knowledge of the Fourier transform and convolution theorem.

\subsection{Problem Statement}
The goal of sequential recommendation is to predict the next item a user will click based on the user's previous interactions.
We define the sequential recommendation issue with the following formulation.
Let $\mathcal{V} = \{v_1, v_2, ..., v_{|\mathcal{V}|}\}$ denote a set of all unique items, and $\mathcal{U} = \{u_1, u_2, ..., u_{|\mathcal{U}|}\}$ denote a set of all users, where $u \in \mathcal{U}$ denotes a user and $v \in \mathcal{V}$ denotes an item. The numbers of users and items are denoted as $|\mathcal{U}|$ and $|\mathcal{V}|$, respectively. The set of user behavior can be represented as $S = \{s_1, s_2,...,s_{|\mathcal{U}|}\}$. In SR, the user's behavior sequence is usually in time order. 
This means that each user sequence is made up of (chronologically ordered) item interactions $s_u = [v_1^{(u)}, v_2^{(u)},..., v_t^{(u)},...,v_{n}^{(u)}]$, 
where $s_u \in S$, $v_t^{(u)} \in \mathcal{V}$ is the item with which user $u$ interacts at step $t$, and $n$ is the length of the sequence. $s_{u,t}=[v_1^{(u)}, v_2^{(u)},...,v_t^{(u)}]$ represents the subsequence of items that the user $u$ interacts with before $t+1$.
Sequences are typically truncated at $N$ maximum length.
Only the most recent $N$ interacting elements are taken into account if the sequence length is larger than $N$.
Zero padding items will be inserted to the left until the sequence length reaches $N$ if it is less than $N$.
As a result, if $t>N$, we truncate the input sequence $s_{u,t}$ to the final $N$ elements when inferring the user $u$ representation at time step $t+1$:
\begin{equation}
    s_{u, t}=\left[v_{t-N+1}, v_{t-N+2}, \ldots, v_{t}\right]
\end{equation}

Specifically, the recommendation model is trained to generate the probability score for each candidate, $i.e.$, $\hat{\mathbf{y}}=\left\{\hat{y}_{1}, \hat{y}_{2}, \ldots, \hat{y}_{|\mathcal{V}|}\right\}$, where $\hat{y_i}$ denotes the prediction score of item $v_i$. 
Given a user's historical interaction sequences without any other auxiliary contextual information, the SR task takes $s_{u,t}$ as input to predict the most possible top-$K$ items at the timestamp $t + 1$, which can be formulated as follows:
\begin{equation}
    v_{u}^{*}=\arg \max _{v_{i} \in \mathcal{V}} P\left(v_{t+1}^{(u)}=v_{i} \mid s_{u,t}\right)
\end{equation}
where $v_{u}^{*}$ represents the groundtruth item of $u$.

\subsection{Fourier Transform}
\label{FT}
Discrete Fourier Transform (DFT)~\cite{DFT1, DFT2} is one of the most widely used computing methods, with numerous applications in data analysis, signal processing, and machine learning. 
In order to transfer sequential behaviors from the time domain to the frequency domain, we introduce DFT into SLIME4Rec.
Since the input data is sequential in SR, we only consider the 1D DFT. Given a finite sequence of $\{x_n\}_{n=0}^{N-1}$, the 1D DFT converts the original sequence into the sequence of complex numbers in the frequency domain by:
\begin{equation}
    X_k=\sum_{n=0}^{N-1} x_n W_N^{nk}, \quad 0 \leq k \leq N-1
\end{equation}
where $W_N^{nk}$ is the twiddle factor. It can be expanded into the following form by Euler's formula:
\begin{equation}
    W_N^{nk} = e^{-\frac{2 \pi i}{N} n k} = \cos \left(\frac{2 \pi}{N} k n\right)-i \cdot \sin \left(\frac{2 \pi}{N} k n\right)
\end{equation}
where $i$ is the imaginary unit of the complex number. $X_k$ is a complex number representing the frequency spectrum of the signal of frequency $\omega_k=2\pi k/N$.
The DFT is obtained by decomposing a sequence of values into components of different frequencies. Note that DFT is a one-to-one unique mapping operation in the time and frequency domains. The sequence of frequency representation $\{X_k\}_{k=0}^{N-1}$ can be transferred to the time feature domain via an Inverse DFT (IDFT), which can be formulated as:
\begin{equation}
    x_{n}=\frac{1}{N} \sum_{k=0}^{N-1} X_{k} W_N^{-nk}
\end{equation}
For real input $x_n$, it has been proven that its DFT is conjugate symmetric, $i.e.$, $X_k=X^*_{N-k}$, where $X^*_{N-k}$ denotes the conjugate of $X_{N-k}$. This property implies that a real discrete signal $x_n$ can be recovered by half of the DFT $ \{X_k\}_{k=0}^{\lceil N / 2\rceil}$~\cite{GFNet}.
Benefited from this property, the Fast Fourier Transform (FFT) algorithm~\cite{cooley, FFTW3} is widely implied to compute DFT, which could reduce the computational complexity from $\mathcal{O}(N^2)$ to $\mathcal{O}(N\log N)$. Consequently, we convert sequential behaviors into the frequency domain via FFT and denote it by $\mathcal{F}(\cdot)$ in our paper. Similar to IDFT, the Inverse FFT (IFFT) (denoted by  $\mathcal{F}^{-1}(\cdot)$) is also used to efficiently transfer the frequency feature back to the time domain.

\subsection{Convolution Theorem}
\label{CT}
Frequency domain convolution has been paid more and more attention in recent years, which opens up another perspective in the field of deep learning.
According to the convolution theorem, 
the circular convolution in the time domain is equivalent to the multiplication in the frequency domain.
This property directly helps us to design filters for capturing different frequency components. In the context of DFT, the convolution theorem can be proved as follows.

Given implicitly periodic signal $x[n]$, the convolution of $x[n]$ and $f[n]$ under this condition is called circular convolution, which can be described by a mathematical formula as:
\begin{equation}
    f[n]* x[n]=\sum_{m=0}^{N-1} f[n] x_{N}[n-m]
\end{equation}
where ``*'' denotes circular convolution and $x_N[\cdot]$ is the period extension of $\{x_n\}_{n=0}^{N-1}$:
\begin{equation}
    x_{N}[n-m] \stackrel{\text { def }}{=} \sum_{k=-\infty}^{\infty} x[n-k N]=\sum_{k=-\infty}^{\infty} x[n+kN]
\end{equation}
It is a special case of circular convolution between two periodic functions that have the same period.
We have the following derivations:
\begin{displaymath} 
\begin{aligned}
 \mathcal{F}(f[n] * x[n]) &= \mathcal{F}(\sum_{m=0}^{N-1} f[m] x_{N}[n-m])\\
    &= \sum_{n=0}^{N-1}\sum_{m=0}^{N-1} f[m] x_{N}[n-m] W_N^{nk} \\
    &= \sum_{u=-m}^{N-1-m} \sum_{m=0}^{N-1} f[m] x_{N}[u] W_N^{(u+m)k} \\
    &= \sum_{m=0}^{N-1} f[m] W_N^{mk} \sum_{u=-m}^{N-1-m} x_{N}[u] W_N^{uk}  \\
    &= \sum_{m=0}^{N-1} f[m] W_N^{mk} \sum_{u=0}^{N-1} x[u] W_N^{uk}  \\
    & = \mathcal{F}(f[n])\cdot \mathcal{F}(x[n])  \\
\end{aligned}
\end{displaymath}
In conclusion, given the transformed frequency feature $X[k]$: 
\begin{equation}
    y[n] =f [n] * x[n] = \mathcal{F}^{-1}(\mathcal{F}(f[n])\cdot X[k])
    \label{convolution}
\end{equation}
In this paper, $\mathcal{F}(f[n])$ is the learnable filter and $y[n]$ is the feature transformed back to the time domain after filtering.
\begin{figure*}[!t]
	\centering
	{\includegraphics[width=1\linewidth]{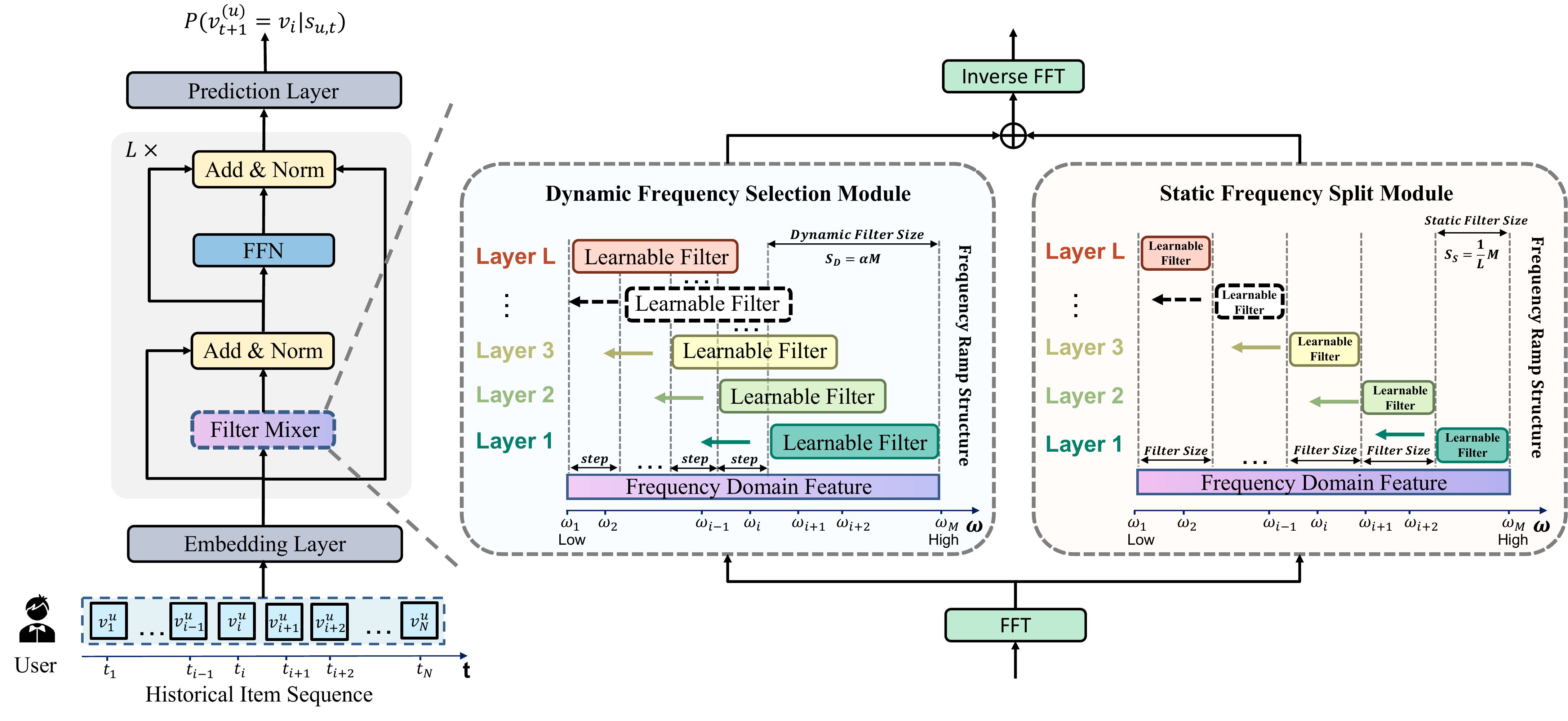}}
    \caption{
    The model architecture of SLIME4Rec is similar to the transformer encoder.
    It first generates item embedding with positional embedding through the embedding layer, then extracts user preference from the frequency domain by the filter mixer layer instead of the self-attention layer.
    The details of the filter mixer are shown on the right side which consists of two modules, $i.e.$, Dynamic Frequency Selection (DFS) and Static Frequency Split (SFS) module.
    For simplicity, we draw all the learnable filters of different layers in one block to show how the frequency ramp structure work on two modules, which looks like the filter slides on the frequency domain feature across different layers.
    Finally, a prediction layer computes a recommendation score for all candidate items.
    }
	\label{model}
\end{figure*}
\section{METHOD}
\label{method}
In this section, we present the contrastive enhanced slide filter mixer architecture (named as \textbf{SLIME4Rec}) for sequential recommendation.
As shown in Figure~\ref{model}, SLIME4Rec is an attention-free Transformer architecture, wherein each block starts with a filter mixer layer and ends with a feed-forward layer.
Essentially, we use the filter mixer layer that consists of Dynamic Frequency Selection (DFS) and Static Frequency Split (SFS) modules to replace the self-attention layer in the transformer encoder.
Then the frequency ramp structure on two modules and the training object are discussed in detail.

\subsection{Embedding Layer}
Sequential recommendation focus on modeling the user behavior sequence of implicit feedback, which is a list of item IDs in SR. 
For all items given in $\mathcal{V}$, we create an item embedding matrix $\mathbf{M}^\mathcal{V}\in \mathbb{R}^{|\mathcal{V}|\times d}$ for all items, where $d$ is the embedding size. User behavior sequence $s_u = [v_1^{(u)}, v_2^{(u)},...,v_{|s_u|}^{(u)}]$ is embedded as:
\begin{equation}
    \mathbf{E}=[\mathbf{e}_{1}^{(u)},\mathbf{e}_{2}^{(u)} , ..., \mathbf{e}_{n}^{(u)}], \mathbf{e}_{k}^{(u)}=\operatorname{LookUp}(v_k^{(u)}, \mathbf{M}^{\mathcal{V}})
\end{equation}
where $\operatorname{LookUp(\cdot, \cdot)}$ retrieves an item embedding from the embedding matrix.

Positional embedding is used to add more positional information while keeping the original embedding dimensions of the item in order to make our model more responsive to the item's position.
We also perform layer normalization and dropout \cite{Dropout} operations to stabilize the training process.
Thus, we generate the sequence representation $\mathbf{E}\in \mathbb{R}^{N\times d}$ by:
\begin{equation}
    \mathbf{E}=\operatorname{Dropout}(\text {LayerNorm}(\mathbf{E}+\mathbf{P}))
\end{equation}

\subsection{Filter Mixer}
To transform the input item representation matrix $\mathbf{E}$ to the frequency domain, we first execute FFT along the item dimension:
\begin{equation}
    \mathbf{H}^{0}=\mathbf{E}
\end{equation}
\begin{equation}
    \mathcal{F} (\mathbf{H}^{l}) \rightarrow \mathbf{X}^{l}\in \mathbb{C}^{M \times d}
\end{equation}
where $\mathbf{H}^{l} \in \mathbb{R}^{N \times d}$ is the time feature of the $l$-th layer, and $\mathcal{F}(\cdot)$ denotes the 1D FFT.
Note that $\mathbf{X}^{l}$ is a complex tensor and represents the spectrum of $\mathbf{H}^{l}$.
As said in section~\ref{FT}, due to the conjugate symmetric property in the frequency domain, half of the DFT contains the full information about the periodic features in the time domain. $M$ is calculated as:
\begin{equation}
    M = {\lceil N / 2\rceil} + 1
\end{equation}

After the Fourier transformation, we get $M$ frequency components $\{\mathbf{X}_k^l\}_{k=0}^{M-1}$ from $\mathbf{X}^{l}\in \mathbb{C}^{M \times d}$.
Frequency $\omega_k=2\pi k/N$ is used to describe the position of these frequency components along the $\omega$-axis in the frequency domain.
To model the frequency-domain features $\mathbf{X}^{l}_k$, we propose filter mixer blocks, where each block consists of two modules, 
$i.e.$, DFS and SFS module, to capture the frequency information in two different ways.

\subsubsection{Dynamic Frequency Selection Module}
In SLIME4Rec, instead of preserving all frequency components, we design a module to dynamically select features at specific frequencies in the frequency domain.
As proved in Section~\ref{CT}, the filter operation in the frequency domain is equal to circular convolution that has a global receptive field like the self-attention mechanism in Transformer architecture and captures the periodic characteristics in the item sequences.
By multiplying a corresponding learnable filter $\mathbf{W}^l_D \in \mathbb{C}^{M \times d}$, we dynamically select the frequency features as:
\begin{equation}
    \label{eq14}
    \widehat{\mathbf{X}}^{l}_D=\mathbf{X}^{l}\odot\sigma^l_{ij}(\omega)\odot\mathbf{W}^l_D
\end{equation}
where $\odot$ is the element-wise multiplication and $l$ indicates that it is the $i$-th layer.
As shown in Figure~\ref{model}, $\mathbf{X}^{l}$ has $M$ frequency components and we use $\omega$ to represent the index of each frequency component.
Therefore, the $\sigma_{ij}^l(\omega)$ is defined as:
\begin{equation}
    \sigma_{i j}^l(\omega)=\mathbb{I}(i^l \leq \omega \leq j^l), \quad 0 \leq i < j \leq M
\end{equation}
Where $\mathbb{I}$ is an indicator function, determined by $i$ and $j$. 
The filter $\mathbf{W}_D^l$ is {\itshape learnable} since it is composed of complex parameters that can be optimized by SGD to adaptively capture sequential pattern characteristics in the frequency domain. 
When the receptive field of the learnable filter in the frequency domain is set as a fixed length window, other frequency components around are shielded to eliminate dependence on irrelevant frequencies.
Therefore, when capturing the characteristics of the current specific frequency range, the learnable filter will simply disable all features that are not in the current range.

\subsubsection{Frequency Ramp Structure} 
As shown in Figure~\ref{model}, to make the filters of each layer cover a specific range, we further design a frequency ramp structure for making the learnable filter slide on the spectrum in different layers. Specifically, for the dynamic frequency selection module, we define $\sigma^l_{ij}(\omega)$ as $\mathop{\sigma^l_{D}(\omega)}\limits ^{\leftarrow}$, where $\leftarrow$ denote the direction of the filter slides, $i.e.$, from high frequency to low frequency ($ \textit{low}\leftarrow\textit{high}$) and $\rightarrow$ denotes the direction from low frequency to high frequency ($\textit{low}\rightarrow\textit{high}$).
It is defined as:
\begin{equation}
    \mathop{\sigma^l_{D}(\omega)}\limits ^{\leftarrow} = \begin{cases}1 & \text { if } \omega  \in[\mathop{i^l_D}\limits ^{\leftarrow}, \mathop{j^l_D}\limits ^{\leftarrow}] \\ 
    0 & \text { otherwise }\end{cases}
\end{equation}
\begin{equation}
    \mathop{i^l_D}\limits ^{\leftarrow} = M * (1 - \alpha) - (l * step)
\end{equation}
\begin{equation}
    \mathop{j^l_D}\limits ^{\leftarrow} = M  - (l * step)
\end{equation}
where $\alpha$ is the ratio of the dynamic filter size $S_D$ relative to the frequency feature range $M$:
\begin{equation}
    \alpha = \frac{S_{D}}{M}
\end{equation}
And given $L$ as the number of filter mixer blocks:
\begin{equation}
    step=\frac{(1-\alpha) M}{L-1}
    \label{step}
\end{equation}
In this way, different layers can capture diverse frequency characteristics. 


When $\alpha$ is set to 1, the dynamic filter size of SLIME4Rec is the same as FMLP-Rec~\cite{FMLP-Rec}, whose the value of Eq.~(\ref{step}) equals 0 and the filter does not slide on the spectrum. For a model with $L$ layers, we get a list of indicator functions of length $L$ which can be denoted as $\mathop{\boldsymbol{\sigma}_{D}(\omega)}\limits ^{\leftarrow}$ = $\{\mathop{\sigma^0_{D}(\omega)}\limits ^{\leftarrow}, \mathop{\sigma^1_{D}(\omega)}\limits ^{\leftarrow}, ...,\mathop{\sigma^{L-1}_{D}(\omega)}\limits ^{\leftarrow}\}$. It can be easily proved that $\mathop{\boldsymbol{\sigma}_{D}(\omega)}\limits ^{\rightarrow}=\operatorname{inverse}(\mathop{\boldsymbol{\sigma}_{D}(\omega)}\limits ^{\leftarrow})$, where $\operatorname{inverse}(\cdot)$ means reverse the order of the input. 
Taking into account that top layers focus more on modeling low-frequency global information while bottom layers are more important for capturing high-frequency details~\cite{iFormer}, we choose the direction from high frequency to low frequency and Eq.~(\ref{eq14}) is rewritten as:
 \begin{equation}
    \widehat{\mathbf{X}}^{l}_D=\mathbf{X}^{l}\odot\mathop{\sigma^l_{D}(\omega)}\limits ^{\leftarrow}\odot\mathbf{W}^l_D
\end{equation}
\subsubsection{Static Frequency Split Module}
The static frequency split module is to make up for the problem of missing some frequency domain features that may exist in the dynamic frequency selection module.
Specifically, when dynamic filter size $S_D < step$, the dynamic filter will not be able to cover the frequency features between the current step and the previous step, $i.e.$: 
\begin{displaymath}
    \begin{aligned}
    \alpha M &<\frac{(1-\alpha) M}{L-1} \\
    (L-1) \alpha &<(1-\alpha) \\
    \alpha &<\frac{1}{L}
    \end{aligned}
\end{displaymath}
So we define the relative ratio of the static filter size as $\beta$:
\begin{equation}
    \beta = \frac{1}{L} = \frac{S_S}{M}
\end{equation}
where $S_S$ is the size of the static filter. Therefore, the slide index $i$ and $j$ of $\mathop{\sigma^l_{S}(\omega)}\limits ^{\leftarrow}$ can be denoted as:
\begin{equation}
    \mathop{i^l_S}\limits ^{\leftarrow} = M * (1 - \beta) - (l * S_S)
\end{equation}
\begin{equation}
    \mathop{j^l_S}\limits ^{\leftarrow} = M  - (l * S_S)
\end{equation}
    
    
    
After that, by multiplying a static learnable filter $\mathbf{W}^l_S \in \mathbb{C}^{M \times d}$, we can modify the spectrum to capture missing frequency features that the dynamic filter cannot capture even if $\alpha<\beta$.
\begin{equation}
    \widehat{\mathbf{X}}^{l}_S=\mathbf{X}^{l}\odot\mathop{\sigma^l_{S}(\omega)}\limits ^{\leftarrow}\odot\mathbf{W}^l_S
\end{equation}
Finally, we mix the features extracted by the dynamic filter and the features extracted by the static filter and adopt the inverse FFT to transform the modulated spectrum $\widehat{\mathbf{X}}^{l}_{mix}$ back to the time domain:
\begin{equation}
    \widehat{\mathbf{X}}^{l}_{mix} = (1 - \gamma) \widehat{\mathbf{X}}^{l}_D + \gamma \widehat{\mathbf{X}}^{l}_S
\end{equation}
\begin{equation}
    \widehat{\mathbf{H}}^{l} \leftarrow \mathcal{F}^{-1} (\widehat{\mathbf{X}}^{l}_{mix}) \in \mathbb{R}^{N \times d}
\end{equation}
where $\mathcal{F}^{-1}$ stands for the inverse 1D FFT, which transforms the complex tensor into a tensor of real numbers.
To address the gradient vanish and unstable training issues, we also include the skip connection, layer normalization, and dropout operations as follows:
\begin{equation}
    \widehat{\mathbf{H}}^{l}=\operatorname{LayerNorm}(\mathbf{H}^{l} + \operatorname{Dropout}(\widehat{\mathbf{H}}^{l}))
\end{equation}

\subsection{Point-wise Feed Forward Network}
Similar to the self-attention mechanism, the frequency-domain convolution in the filter mixer is still a linear operation, which fails to model complex non-linear relations. 
To endows the models with non-linearity characteristics between different dimensions in the time domain, we also add a feed-forward network after each filter mixer, which consists of Multi-Layer Perceptron (MLP) with GELU activation.
The process of the point-wise Feed-Forward Neural network (FFN) is defined as follows: 
\begin{equation}
    \tilde{\mathbf{H}}^{l}=\operatorname{FFN}(\widehat{\mathbf{H}}^{l})=(\operatorname{GELU}(\widehat{\mathbf{H}}^{l} \mathbf{W}_{1}+\mathbf{b}_{1})) \mathbf{W}_{2}+\mathbf{b}_{2}
\end{equation}
where $\mathbf{W}_{1}, \mathbf{W}_{2} \in \mathbb{R}^{d\times d}$ and $\mathbf{b}_{1}, \mathbf{b}_{2} \in \mathbb{R}^{1\times d}$ are learnable parameters. 
In order to prevent overfitting, we add a dropout layer above each hidden layer and perform layer normalization procedures again using densely residual connection structure on the output $\mathbf{H}^{l+1}$, as below:
\begin{equation}
    \mathbf{H}^{l+1}=\operatorname{LayerNorm}(\mathbf{H}^{l} + \widehat{\mathbf{H}}^{l} + \operatorname{Dropout}(\tilde{\mathbf{H}}^{l}))
\end{equation}

\subsection{Prediction Layer}
After $L$ filter mixer blocks that adaptively and hierarchically extract behavior pattern information of previously interacted items, we get the final combined representation of items that represent the user preference.
Based on the $L$-layer encoder output $\mathbf{H}^{L}$, we select the last hidden vector $\mathbf{h}^L_t$ in $\mathbf{H}^{L}=[\mathbf{h}^L_0, \mathbf{h}^L_1, \cdots, \mathbf{h}^L_t]$ as the user representation of this user sequence, we can compute the recommendation probability of candidate item $v$ to predict how likely the user would adopt the item. 
Specifically, the corresponding predicted probability $\mathbf{\hat{y}}$ can be generated by:
\begin{equation}
    \mathbf{{\hat{y}}}=\operatorname{softmax}((\mathbf{M}^\mathcal{V})^{\top}\mathbf{h}^{L})
\end{equation}
where $\mathbf{\hat{y}}\in \mathbb{R}^{|\mathcal{V}|}$.
As a result, we expect that the true item $v$ that user $u$ adopted will lead to a higher score $\hat{y_i}$.
To optimize the model parameter, we therefore use the cross-entropy loss. 
The objective function of SR can be formulated as:
\begin{equation}
    \mathcal{L}_{Rec}=-\sum_{i=1}^{|\mathcal{V}|} y_{i} \log \left(\hat{y}_{i}\right)+\left(1-y_{i}\right) \log \left(1-\hat{y}_{i}\right)
    \label{LRec}
\end{equation}

\subsection{Contrastive Learning}
To enhance the training of the filters in both dynamic frequency selection and static frequency split modules to capture the core frequency components of user sequence, we leverage a multi-task training strategy to jointly optimize the main recommendation loss with auxiliary contrastive loss. It helps to minimize the difference between differently augmented views of the same user and maximize the difference between the augmented sequences derived from different users.

Although previous augmentations methods~\cite{CL4SRec} including item cropping, masking, and reordering help to enhance the performance of SR models, the data-level augmentations cannot 
guarantee a high level of semantic similarity~\cite{DuoRec}.
Instead of using typical data augmentations, we let the same user’s sequence pass through the network twice and model the frequency components to construct harder positive samples by mixing the frequency feature extract from DFS and SFS before the inverse Fourier transform.
Since there are different dropout layers in the network module, we will get two output views with different numerical features but similar semantics.
Besides, in order to increase the supervision signal of contrast learning, we follow DuoRec~\cite{DuoRec} to use a sequence with the same target as a positive sample of supervised contrast learning.
All the other augmented samples in the training batch are treated as negative samples in order to efficiently create the negative samples for an augmented pair of samples (represented as $neg$).

For the batch $\mathcal{B}$, the contrastive regularization is defined as:
\begin{equation}
    \mathcal{L}_{\mathrm{CLReg}}=\mathcal{L}_{\mathrm{CLReg}}(\mathbf{h}^{\prime}, \mathbf{h}_s^{\prime})+\mathcal{L}_{\mathrm{CLReg}}(\mathbf{h}_s^{\prime}, \mathbf{h}^{\prime})
\end{equation}
\begin{equation}
    \mathcal{L}_{\text {CLReg}}(\mathbf{h}^{\prime}, \mathbf{h}_s^{\prime})=-\log \frac{\exp (\operatorname{sim}(\mathbf{h}^{\prime}, \mathbf{h}_s^{\prime}))}{\sum_{n e g} \exp (\operatorname{sim}(\mathbf{h}^{\prime}, \mathbf{h}_{neg}))}
\end{equation}
where $\mathbf{h}^{\prime}$ and $\mathbf{h}_s^{\prime}$ represent unsupervised and supervised augmented views, respectively, defined as follows:
\begin{equation}
    \mathbf{h}^{\prime}=\mathbf{H}^{L^{\prime}}[-1], \quad \mathbf{h}_s^{\prime}=\mathbf{H}_s^{L^{\prime}}[-1]
\end{equation}
Thus, the overall objective of SLIME4Rec is:
\begin{equation}
    \ell = \ell_{\mathrm{Rec}} + \lambda\ell_{\mathrm{CLReg}}
\end{equation}
where $\lambda$ is hyperparameter to control the strengths of contrastive regularization.
\subsection{Complexity Analysis}
Different from transformer-based models that rely on the self-attention mechanism, SLIME4Rec is an attention-free architecture with dynamic and static filters.
The computation complexity of traditional self-attention is $\mathcal{O}(n^2d+nd^2)$, where $n$ is the sequence length of user and $d$ is hidden size.
In contrast, the time complexity of the filter mixer block can be reduced to $\mathcal{O}(n\log(nd))$ with FFT~\cite{conjugate}. 
The time complexity of the element-wise product in our proposed dynamic frequency selection and static frequency split module is $\mathcal{O}(nd)$, and the time cost of feed-forward networks is $\mathcal{O}(\frac{nd^2}{2})$.
Therefore, considering contrastive objectives and layer numbers the total time complexity of SLIME4Rec is $\mathcal{O}(3L(n\log{n}d + nd + \frac{nd^2}{2}))$ which is proportional to the log-linear complexity of the input sequence length $n$.

\section{EXPERIMENTS}
In this section, we first briefly introduce the datasets used in our experiments, eight baselines, the evaluation metrics, and the implementation details in our experimental settings.
Then, we compare our proposed model SLIME4Rec with state-of-the-art baseline methods, 
present the experimental results of each model and analyze the reasons.
Additionally, we explore how the performance of our model SLIEM4Rec is impacted by important model components and parameters.
Finally, we discuss the robustness of SLIME4Rec to synthetic noises and present the visualization
of dynamic and static filters. 
Specifically, to study the validity of SLIME4Rec, we conduct experiments to try to answer the following questions:
\begin{itemize}
    \item {\bfseries RQ1} 
    Does SLIME4Rec perform better than the state-of-the-art baselines?
    \item {\bfseries RQ2} What is the influence of dynamic and static filter modules in the SLIME4Rec?
    \item {\bfseries RQ3} What is the influence of filter slide direction on model performance?
    \item {\bfseries RQ4} How do the parameters, such as dynamic filter size ratio $\alpha$, max item length $N$, hidden size $d$, and the number of filter mixer blocks $L$, affect the effectiveness of SLIME4Rec?
    \item {\bfseries RQ5} How robust is the proposed model to synthetic noise?
\end{itemize}
\subsection{Dataset}
We conduct experiments on five public datasets collected from real-world platforms in order to thoroughly evaluate SLIME4Rec.
These datasets, which differ in scenarios, sizes, and sparsity, are frequently used in tests of sequential recommendation methods.
The main statistics of five datasets after preprocessing are reported in Table~\ref{tab:seq1}.
We elaborate on the descriptions of the individual dataset below.
\begin{itemize}
\item {\bfseries Beauty, Clothing, and Sports}~\cite{Amazon} 
are three datasets that were gathered from Amazon, one of the biggest e-commerce platforms in the world.
They are divided by the highest-level product categories on Amazon.
High sparsity and a variety of rating review categories are characteristics of Amazon datasets.
We follow the setting in DuoRec and adopt three categories, ``Beauty", ``Clothing Shoes and Jewelry", and ``Sports and Outdoors".

\item {\bfseries MovieLens-1M}~\cite{ML-1M} 
is based on reviews of movies that were collected from the non-commercial movie recommendation website MovieLens.
The interaction number in \emph{ML-1M} is about 1 million.

\item {\bfseries Yelp}~\cite{DuoRec} is a famous dataset for business recommendation. 
We only use the transaction records from after January 1st, 2019, due to the scale of transactions.
\end{itemize}
Following \cite{Timeinterval, BERT4Rec}, we also adopt the 5-core settings by filtering out users with less than 5 interactions. 

\subsection{Evaluation Metrics}
In our evaluation, we adopt the leave-one-out strategy, in which the final item that a user interacts with is held out for testing, the next-to-last item is held out for validation, and the remaining items are held out for training.
We rank the prediction scores throughout the entire item set without using negative sampling, as recommended by~\cite{kdd}, which ensures that the evaluation process is unbiased.
Performance is evaluated on a variety of evaluation metrics, including Hit Ratio at $K$ (HR@$K$) and Normalized Discounted Cumulative Gain at $K$ (NDCG@$K$) on all datasets. Note that higher values of both indicate better performance. 
HR@$K$ is a recall-based metric that measures the average proportion of ground-truth items in the top-$K$ recommendation lists. 
NDCG@$K$ is a position-aware metric that evaluates the ranking quality of the top-$K$ recommendation lists in a position-wise manner. 
We report HR and NDCG for $K$ = 5 and 10 in this paper.
\begin{table}[t]
\renewcommand{\arraystretch}{1.2}
    \centering
    \caption{Statistics of the datasets after preprocessing.}
    \label{tab:seq1}
    \begin{adjustbox}{max width=\linewidth}
        \begin{tabular}{l| r r r r r}
        \toprule
        Specs. & Beauty & Clothing & Sports & ML-1M & Yelp\\
        \midrule
        \midrule
       \# Users & 22,363 & 39,387 & 35,598 &  6,041 & 30,499 \\
       \# Items & 12,101 & 23,033 & 18,357 & 3,417 & 20,068 \\
       \# Avg.Length & 8.9 & 7.1 & 8.3 & 165.5 & 10.4\\
       \# Actions & 198,502 & 278,677 & 296,337 & 999,611 & 317,182\\
       Sparsity & 99.93\% & 99.97\% & 99.95\% & 95.16\% & 99.95\%\\
       \bottomrule
    \end{tabular}
    \end{adjustbox}
    \label{dataset}
\end{table}
\subsection{Baseline Models}
To demonstrate the effectiveness of the proposed model, we compare SLIME4Rec with the most wide-used and state-of-the-art methods, including plain matrix factorization methods (BPR-MF), sequential models with different neural architectures (GRU4Rec, Caser, SASRec, BERT4Rec, FMLP-Rec), and competitive sequential models using self-supervised learning contrastive objective function (CL4SRec, ContrastVAE, CoSeRec, and DuoRec). All baselines are described as follows:

\textbf{BPR-MF}\cite{BPR_MF} is a classic non-sequential method for learning personalized ranking from implicit feedback and optimizing the matrix factorization through a pair-wise Bayesian Personalized Ranking (BPR) loss.

\textbf{GRU4Rec}\cite{GRU4Rec} is the first model to apply Gated Recurrent Unit (GRU) to model sequences of user behavior for sequential recommendation. 
The final hidden feature of the GRU is treated as the latent representation of the input sequence.

\textbf{Caser}\cite{Caser} is a CNN-based method capturing local dynamic patterns of user activity by using horizontal and vertical convolutional filters over time.

\textbf{SASRec}\cite{SASRec} 
is one of the Transformer-based models with the multi-head self-attention mechanism. 
Since the powerful capability to model long-term dependence, SASRec has become a strong baseline in the sequential recommendation. 

\textbf{BERT4Rec}\cite{BERT4Rec} 
uses a masked item traning scheme similar to the masked language model sequential in NLP.
The backbone is the bi-directional self-attention mechanism.

\textbf{FMLP-Rec}\cite{FMLP-Rec} is a all-MLP model using a learnable filter-enchanced block to remove noise in the embedding matrix.
Compared to traditional sequence modeling techniques, this structure is significantly different.

\textbf{CL4SRec}\cite{CL4SRec}
uses three data augmentation techniques, including sequence cropping, masking, and reordering, to generate different contrastive representations of the same user interaction sequence for the auxiliary contrastive learning task.

\textbf{ContrastVAE}\cite{ContrastVAE} 
is a VAE-based method that combines three augmentations strategies including data augmentation, model augmentation, and variational augmentation for contrastive learning.

\begin{table*}
\renewcommand\arraystretch{1.1}
\setlength{\tabcolsep}{0.6em}
    \centering
    \caption{
    Overall performance.
    The Bold scores indicate the best results of all methods.
    Underlined scores stand for the suboptimal results of all methods.
    The SLIME4Rec outperforms all baseline models by a large margin in terms of all evaluation metrics.
    } %
    \begin{adjustbox}{max width=\textwidth}
        \begin{tabular}{l|l|c| c c c c c |c c c c|c|c}

        \toprule
        Datasets & Metric & BPR-MF & GRU4Rec & Caser & SASRec & BERT4Rec & FMLP-Rec & CL4SRec &ContrastVAE & CoSeRec & DuoRec & SLIME4Rec & Improv.\\
        
        \midrule
        \multirow{4}{*}{Beauty} 
            &HR@5	&0.0120	&0.0164	&0.0259	&0.0365 &0.0193	&0.0398	&0.0401 &0.0422	&0.0537	&\underline{0.0546}	&\textbf{0.0621}	&13.74\%\\
            &HR@10	&0.0299	&0.0365	&0.0418	&0.0627 &0.0401	&0.0632	&0.0683 &0.0681	&0.0752	&\underline{0.0845}	&\textbf{0.0910}	&7.69\%\\
            &NDCG@5	&0.0040	&0.0086	&0.0127	&0.0236 &0.0187	&0.0258	&0.0223 &0.0268	&0.0361	&\underline{0.0352}	&\textbf{0.0396}	&12.50\%\\
            &NDCG@10	&0.0053	&0.0142	&0.0253	&0.0281 &0.0254	&0.0333	&0.0317 &0.0350	&0.0430	&\underline{0.0443}	&\textbf{0.0489}	&10.38\%\\
            

        
        \midrule
        \multirow{4}{*}{Clothing} 
            &HR@5	&0.0067	&0.0095	&0.0108	&0.0168 &0.0125	&0.0126	&0.0168 &0.0161	&0.0175	&\underline{0.0193}	&\textbf{0.0225}	&16.58\%\\
            &HR@10	&0.0094	&0.0165	&0.0174	&0.0272	&0.0208 &0.0206 &0.0266 &0.0247	&0.0279	&\underline{0.0302}	&\textbf{0.0343}	&13.58\%\\

            &NDCG@5	&0.0052	&0.0061	&0.0067	&0.0091	&0.0075 &0.0082	&0.0090 &0.0105	&0.0095	&\underline{0.0113}	&\textbf{0.0126}	&11.50\%\\
            &NDCG@10 &0.0069	&0.0083	&0.0098	&0.0124 &0.0102	&0.0107	&0.0121 &0.0133	&0.0131	&\underline{0.0148}	&\textbf{0.0164}	&10.81\%\\
         
        \midrule
        \multirow{4}{*}{Sports}
            &HR@5	&0.0092	&0.0137	&0.0139	&0.0218	&0.0176 &0.0218	&0.0227 &0.0225	&0.0287	&\underline{0.0326}	&\textbf{0.0373}	&14.42\%\\
            &HR@10	&0.0188	&0.0274	&0.0231	&0.0336	&0.0326 &0.0344	&0.0374 &0.0366	&0.0437	&\underline{0.0498}	&\textbf{0.0565}	&13.45\%\\
            &NDCG@5	&0.0040	&0.0096	&0.0085	&0.0127	&0.0105 &0.0144	&0.0129 &0.0151	&0.0196	&\underline{0.0208}	&\textbf{0.0243}	&16.83\%\\
            &NDCG@10	&0.0051	&0.0137	&0.0126	&0.0169	&0.0153 &0.0185	&0.0197 &0.0184	&0.0242	&\underline{0.0262}	&\textbf{0.0305}	&16.41\%\\

        \midrule
        \multirow{4}{*}{ML-1M} 
            &HR@5	&0.0078	&0.0763	&0.0816	&0.1087 &0.0733	&0.1356 &0.1147 &0.1406	&0.1262	&\underline{0.2038}	&\textbf{0.2237}	&9.76\%\\
            &HR@10	&0.0162	&0.1658	&0.1593	&0.1904	&0.1323 &0.2118	&0.1975 &0.2220	&0.2212	&\underline{0.2946}	&\textbf{0.3156}	&7.13\%\\
            &NDCG@5	&0.0052	&0.0385	&0.0372	&0.0638 &0.0432	&0.0870	&0.0662 &0.0895	&0.0761	&\underline{0.1390}	&\textbf{0.1567}	&12.73\%\\
            &NDCG@10 &0.0079 &0.0671	&0.0624	&0.0910 &0.0619	&0.1113 &0.0928  &0.1157	&0.1021	&\underline{0.1680}	&\textbf{0.1864}	&10.95\%\\
            
        \midrule
        \multirow{4}{*}{Yelp} 
            &HR@5	&0.0127	&0.0152	&0.0156	&0.0161 &0.0186	&0.0179	&0.0216 &0.0177	&0.0241	&\underline{0.0441}	&\textbf{0.0516}	&17.01\%\\
            &HR@10	&0.0245	&0.0263	&0.0252	&0.0265	&0.0291 &0.0304	&0.0352 &0.0294	&0.0395	&\underline{0.0631}	&\textbf{0.0766}	&21.39\%\\
            &NDCG@5	&0.0760	&0.0104	&0.0096	&0.0102 &0.0118	&0.0113	&0.0130 &0.0113	&0.0151	&\underline{0.0325}	&\textbf{0.0359}	&10.46\%\\
            &NDCG@10	&0.0119	&0.0137	&0.0129	&0.0134	&0.0171 &0.0153	&0.0185 &0.0147	&0.0205	&\underline{0.0386}	&\textbf{0.0439}	&13.73\%\\
        \midrule
        \bottomrule
        \end{tabular}
    \end{adjustbox}
    \label{tab:result}
\end{table*}

\textbf{CoSeRec}\cite{CoSeRec} 
makes use of correlations between items to improve the robustness of data augmentation within the contrastive framework.

\textbf{DuoRec}\cite{DuoRec} uses unsupervised model-level augmentation and supervised semantic positive samples for contrastive learning. 
It is the most recent and strong baseline for sequential recommendation.

\subsection{Implementation Details}

We implement our SLIME4Rec model in PyTorch.
All the experiments are conducted on a Linux server equipped with a 32GB NVIDIA Tesla V100 GPU. 
For the baseline models, we refer to their best hyper-parameters setups reported in the original papers and directly report their reimplementations results if available, since the datasets and evaluation metrics used in these works are strictly consistent with ours.

The model is optimized by Adam optimizer with a learning rate of 0.001.
Both the dimension of the feed-forward network and item embedding size are set to 64.
For the dropout rate on the embedding matrix and filter mixer blocks are chosen from \{0.1, 0.2, 0.3, 0.4, 0.5\}.
Thanks to the frequency ramp structure, we stack more filter layers and search the total number of filter mixer layers in \{2, 4, 8\}.
Due to the $\mathcal{O}(N\log N)$ complexity, the maximum sequence length $N$ can be chosen from \{25, 50, 75, 100\} without adding too much computation cost. 
For the dynamic frequency selection modules, we set the $\alpha$ as hyper-parameters and select it from [0, 1] with step 0.1.
All these parameters are tuned on the validation set.
We report the result of each model under its optimal hyper-parameter settings.
The implementation of our model can be found at https://github.com/sudaada/SLIME4Rec.


\subsection{Recommendation Performance Comparison}
To prove the sequential recommendation performance of our model SLIME4Rec, we compare it with other state-of-the-art methods (\textbf{RQ1}).
Table~\ref{tab:result} presents the detailed evaluation results of each model where the results of our SLIME4Rec and the strongest baselines are highlighted in bold and underlined respectively. According to the results, we can draw the following observations and conclusions.

First, it is no doubt that the non-sequential recommendation method BPR-MF 
displays the lowest results across all datasets since it ignores the sequential information, 
which indicates the importance of mining the sequential patterns hidden inside user behavior sequences.
Second, all of the neural network methods (GRU4Rec, Caser, SASRec, BERT4Rec, and FMLP-Rec) are significantly better than conventional methods, proving the neural network based models are capable of capturing complex sequential patterns for making recommendations.
For example, GRU4Rec performs better than BPR-MF by leveraging the recurrent structure to capture the user's general preference. 
We can draw the conclusion that using sequential information can enhance performance.
Moreover, Caser achieves better performance than GRU4Rec on most datasets, which indicates the effectiveness of convolution kernels to capture the more complex behavior pattern. Compared with the previous RNN-based and CNN-based methods, the advanced Transformer-based method (e.g., SASRec) improves the performance by a large margin. It demonstrates that the self-attention mechanism has a stronger capability of modeling interaction sequences in SR. 
For example, BERT4Rec applies the masked item prediction objective.
Although such a task can introduce a meaningful signal for the model, the performance is not consistent since the masked item prediction is not aligned well with the recommendation task.
More recently, a filter-enhanced MLP structure achieved almost the same performance as SASRec or even better on most datasets by attenuating the noise in the frequency domain.

Third, compared with the vanilla method, the model with auxiliary self-supervised learning tasks gains decent improvement. 
For example, CL4SRec improves the performance of SASRec with three data augmentation.
ContrastVAE apply variational augmentations to create a stronger augmented view for contrastive learning, which achieves comparable or even better results than CL4SRec.
CoSeRec follows the pipeline of CL4SRec to enhance contrast sequences by leveraging item correlations. 
It demonstrates how contrastive learning could be a useful method for sequential recommendation tasks.
DuoRec outperforms all the baselines by a large margin by model augmentation and semantic augmentation, which verify the effectiveness of the combination of supervised contrastive learning and unsupervised contrastive learning.

Finally, our model outperforms other competing methods on both sparse and dense datasets with a significant margin across all the metrics, demonstrating the superiority of our model. 
Specifically, SLIME4Rec achieves remarkable improvements over the strongest baselines w.r.t. NDCG@5 improves by 16.58\% and 12.73\% on Amazon Sports and ML-1M, respectively, and HR@10 improves by 21.39\% on Yelp.
This observation demonstrates the effectiveness of SLIME4Rec and shows that transforming user sequence from the time domain to the frequency domain and multiplying a learnable filter of the appropriate size on the frequency components is promising to extract useful information for accurate recommendation.
\begin{figure}[!t]
	\centering
	{\includegraphics[width=1\linewidth]{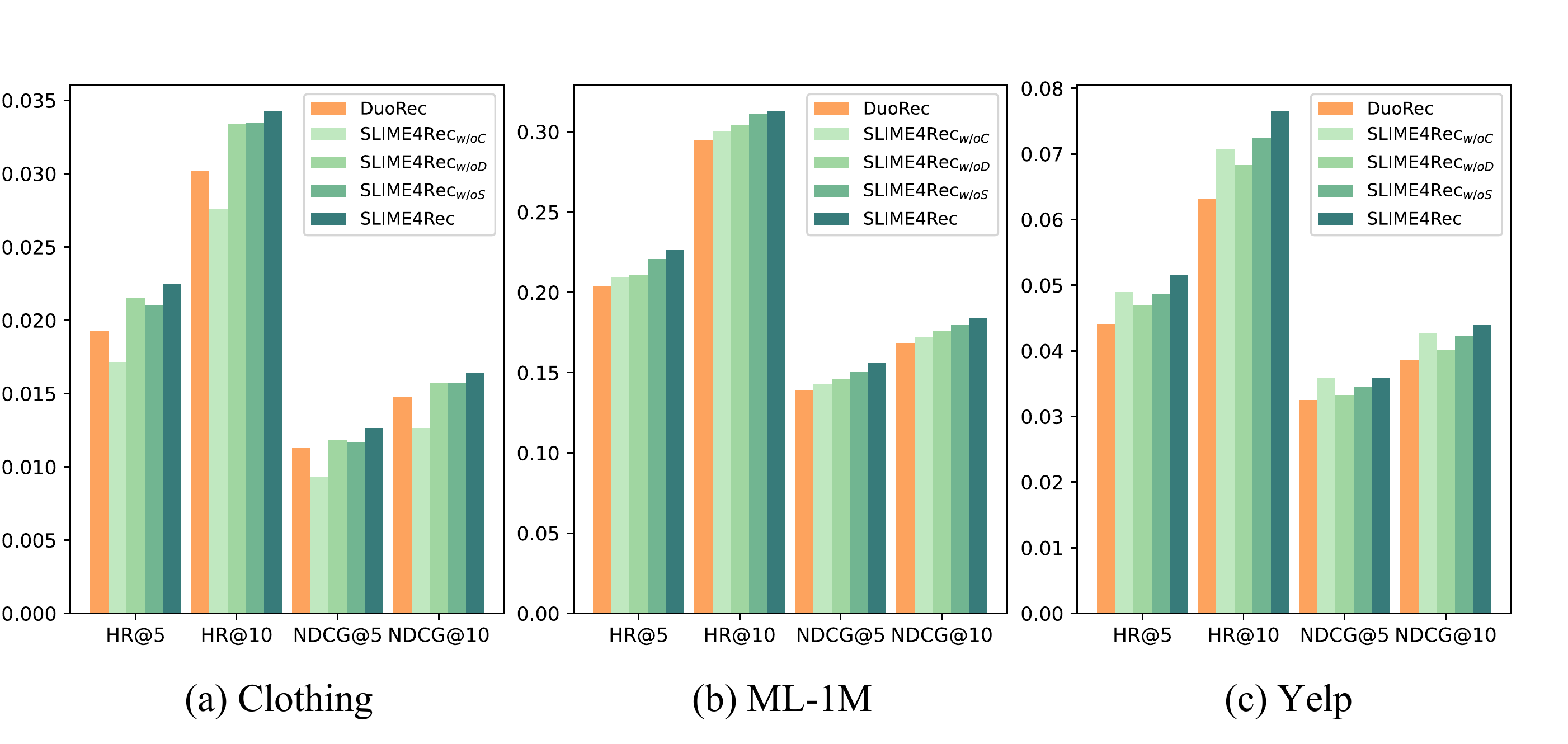}}
	\caption{
        Ablation study of the different filter module and contrastive learning.
        }
	\label{fig:ablation}
\end{figure}

\subsection{Ablation Study}
As introduced in Section III, the backbone architecture of SLIME4Rec is constructed with two main novel components, $i.e.$, the dynamic and static filter module and the frequency ramp structure. To better understand the design rationale of our method, we conduct some ablations study on the proposed model to investigate the effectiveness of these components separately (\textbf{RQ2}). 
Due to the space limit, we only show the analysis results in terms of HR@5 and NDCG@5. 
We have obtained similar experimental results in terms of other metrics. 

            
            
            
            
            
            
\begin{table}[h]\Huge
\renewcommand{\arraystretch}{1.2}
    \centering
    \caption{
    Performance (HR@5 and NDCG@5) of the different filter module designs.
    }
    \label{tab:seq3}
    \begin{adjustbox}{max width=\linewidth}
        \begin{tabular}{c|cc|rc|rc|rc|rc|rc}
        \hline
        \hline
        \multicolumn{1}{c|}{
        \multirow{2}{*}{Layer}
        }
    
        & \multirow{2}{*}{DFS} 
        & \multirow{2}{*}{SFS} 
        & \multicolumn{2}{c|}{Beauty}  
        & \multicolumn{2}{c|}{Clothing} 
        & \multicolumn{2}{c|}{Sports} 
        & \multicolumn{2}{c|}{ML-1M} 
        & \multicolumn{2}{|c}{Yelp}\\
        \cline{4-13}
        
        \multicolumn{1}{c|}{}
        &
        &
        & HR@5
        & NDCG@5
        & HR@5
        & NDCG@5
        & HR@5
        & NDCG@5
        & HR@5
        & NDCG@5
        & HR@5
        & NDCG@5\\
        \hline
        \hline
    
        \multirow{2}{*}{L=2}
        &$\alpha$=0.3 &\xmark	&0.0588 &0.0360	&0.0209 &0.0116	&0.0357 &0.0227 &0.1876 &0.1287 &0.0449 &0.0317\\
        &$\alpha$=0.3 & $\beta$=0.5	&\textbf{0.0604} &\textbf{0.0370}	&\textbf{0.0210} &\textbf{0.0118}	&\textbf{0.0358} &\textbf{0.0228} &\textbf{0.1907} &\textbf{0.1312} &\textbf{0.0454} &\textbf{0.0320} \\
        \hline
        
        \multirow{2}{*}{L=4}
        &$\alpha$=0.2 &\xmark &0.0594 &0.0373 &0.0213 &0.0121 &0.0367 &0.0234 &0.1874 &0.1273 &0.0467 &0.0327\\
        &$\alpha$=0.2 &$\beta$=0.25&\textbf{0.0599} &\textbf{0.0376}	&\textbf{0.0217} &\textbf{0.0124} &\textbf{0.0369} &\textbf{0.0235} &\textbf{0.1879} &\textbf{0.1274} &\textbf{0.0481} &\textbf{0.0337}\\
        \hline
        
        \multirow{2}{*}{L=8}
        &$\alpha$=0.1 &\xmark	&0.0570 &0.0371 &0.0203 &0.0120	&0.0365 &0.0232 &0.1945 &0.1357 &0.0452 &0.0312\\
        &$\alpha$=0.1 &$\beta$=0.125	&\textbf{0.0591} &\textbf{0.0379}	&\textbf{0.0211} &\textbf{0.0128}	&\textbf{0.0369} &\textbf{0.0239}	&\textbf{0.2020} &\textbf{0.1384} &\textbf{0.0460} &\textbf{0.0327}\\
       \hline
       \hline
    \end{tabular}
    \end{adjustbox}
\end{table}



\subsubsection{Investigation of Filter Module Design}
We start by exploring the influence of different filter module designs and how contrastive learning enhance these modules.
To investigate the contribution of each component to the final recommendation performance, we design three variants of SLIME4Rec.
When the contrastive learning, dynamic filter or static filter is removed from SLIME4Rec, the model is degraded to SLIME4Rec$_{w/oC}$, SLIME4Rec$_{w/oD}$ and SLIME4Rec$_{w/oS}$ respectively.
SLIME4Rec has both dynamic and static filter modules, which is the full version of our proposed model.

Specifically, we conduct an ablation study on three datasets and show the results in Figure~\ref{fig:ablation}. 
From the results, we can observe that removing any component would lead to performance degradation and all the variants of SLIME4Rec perform better than DuoRec, proving that all components of the proposed model are effective and necessary.
Besides, we can see that SLIME4Rec outperforms all the other variants with a single filter module. 
The reason is that dynamic and static feature mix can learn the sequential pattern from different receptive fields jointly, which is promising to better capture user preference. 
\begin{figure*}[!t]
	\centering
	{\includegraphics[width=1\linewidth]{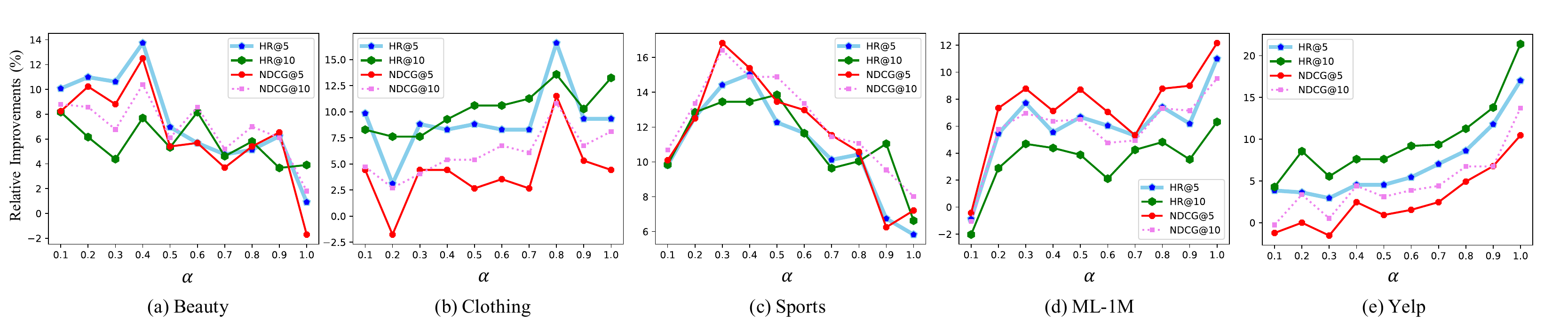}}
	\caption{Relative improvement with different dynamic filter sizes $\alpha$ compared to the strongest baseline DuoRec.}
	\label{fig:alpha}
\end{figure*}

To prove the necessity of using a static filter module, we conduct a detailed ablation study when the dynamic filter size is too small to capture the frequency range of the current step.
We analyze the importance of local awareness in Table~\ref{tab:seq3}.
Table~\ref{tab:seq3} shows that although the dynamic filter module plays a more critical role in capturing frequency features, the static filter plays an important supporting role when $\alpha<\beta$.
Benefiting from both using dynamic and static filter modules, SLIME4Rec can preserve complete information on spectrum characteristics in each layer when $\alpha < \beta$ and collect all informative signals across different layers without losing information of any frequency range.
        
        
        
\begin{table}[h]\Huge
\renewcommand{\arraystretch}{1.4}
    \centering
    \caption{
    Performance(HR@5 and NDCG@5) of the different filter slide modes of frequency ramp structure.
    }
    \label{tab:seq4}
    \begin{adjustbox}{max width=\linewidth}
        \begin{tabular}{l|cc|rc|rc|rc|rc|rc}
        \hline
        \hline
        \multicolumn{1}{c|}{
        \multirow{2}{*}{Slide}
        }
        & \multirow{2}{*}{DFS} 
        & \multirow{2}{*}{SFS} 
        & \multicolumn{2}{c|}{Beauty}  
        & \multicolumn{2}{c|}{Clothing} 
        & \multicolumn{2}{c|}{Sports} 
        & \multicolumn{2}{c|}{ML-1M} 
        & \multicolumn{2}{c}{Yelp}\\
        \cline{4-13}
        \multicolumn{1}{c|}{}
        &
        &
        & HR@5
        & NDCG@5
        & HR@5
        & NDCG@5
        & HR@5
        & NDCG@5
        & HR@5
        & NDCG@5
        & HR@5
        & NDCG@5 \\
        \hline
        \hline
        
        Mode 1
        &$\leftarrow$ &$\rightarrow$	&0.0577 &0.0371	&0.0216	&0.0120
 &0.0360 &0.0239
	&0.2086 &0.1432 &0.0486 &0.0343
\\
        
        Mode 2
        &$\rightarrow$ &$\leftarrow$
        &0.0563 &0.0360
	&0.0214	&0.0121
 &0.0361 &0.0224
 &0.2104 &0.1461
 &0.0489 &0.0346
\\
        
        Mode 3
        &$\rightarrow$ &$\rightarrow$	
        &0.0589 &0.0371
	&0.0220 &0.0123
	&0.0367 &0.0233
	&0.2108 &0.1455 &0.0493 &0.0343
\\
        
        Mode 4
        &$\leftarrow$ &$\leftarrow$	&\textbf{0.0621} &\textbf{0.0396} &\textbf{0.0225} &\textbf{0.0126}	&\textbf{0.0373} &\textbf{0.0243} &\textbf{0.2237} &\textbf{0.1567} 
        &\textbf{0.0516} &\textbf{0.0359}\\
       \hline
       \hline
    \end{tabular}
    \end{adjustbox}
\end{table}

\subsubsection{Investigation of Frequency Ramp Structure}
SLIME4Rec mainly focuses on extracting features in the frequency domain, and one of its most innovative structures is the sliding filter.
Since the features of low frequency and high frequency are distributed on the left and right sides respectively in the frequency domain, there are two sliding directions.
In both the dynamic module and static module, we adopt the slide technique to fuse information about users' behavior patterns with different frequencies from the feature in the frequency domain. 
To further illustrate the impact of different slide modes on SLIME4Rec, we conduct ablation studies on five datasets (\textbf{RQ3}).
As shown in Table~\ref{tab:seq4}, there are four slide modes. Symbol $\leftarrow$ stands for the filter sliding from high-frequency feature to low-frequency feature. Symbol $\rightarrow$ stands for the filter sliding from low-frequency feature to high-frequency feature.

We note that sliding mode 4 outperforms other sliding modes.
This implies that the details in the high-frequency feature are suitable for the lower layers filter to capture the local frequently changed interest of the user. As the network layer goes deeper, various local information is gradually collected to understand the entire global interaction sequence.
Besides, mode 3 achieves the second-best performance.
The sliding mode 3 is very similar to mode 4, but in opposite directions.
The dynamic and static filter in mode 3 captures low-frequency feature at the bottom layer and high-frequency information at the top layer, which is inconsistent with the conclusion in \cite{iFormer} that lower layers often need more local information, while higher layers desire more global information.

In fact, the dynamic and static filters in modes 1 and 2 have opposite and conflicting sliding directions no matter whether they are on the top layer or the bottom layer.
In other words, the low-frequency information extracted by the dynamic filter is directly mixed with the high-frequency information extracted by the static filter, and vice versa.
Therefore adding these two conflicts frequency feature  leads to suboptimal results.
Applying frequency ramp structure on the filter module makes SLIME4Rec better trade-off high- and low-frequency, since the layers at different positions in the neural network play different roles in capturing feature details. 
SLIME4Rec further models sequential patterns from high-frequency to low-frequency in the frequency ramp structure, which is suitable for feature extraction of deep neural networks \cite{iFormer}.

\subsection{Influence of Hyper-parameters}
To explore the effectiveness of capturing sequential patterns in the frequency domain with a sliding filter, we study how four hyperparameters, the filter size ratio $\alpha$, the model depth $L$, hidden size $d$, and the maximum length of user sequence $M$, affect the performance of SLIME4Rec (\textbf{RQ4}).
We analyze one hyper-parameter at a time by fixing the remaining hyper-parameters at their optimal settings.


\subsubsection{Impact of Filter Size Ratio $\alpha$}
One of the novelties in the proposed SLIME4Rec model is that we let a filter of appropriate size slide over the frequency domain features across different layers.
So in our work, we introduce the hyperparameter $\alpha$ to control the filter size, which can significantly affect the performance of our model.
We conduct experiments under different $\alpha$ on five datasets and illustrate  the impact of this hyperparameter in Figure~\ref{fig:alpha}.

As shown in Figure~\ref{fig:alpha},  we can observe that with the increase of $\alpha$ the performance of SLIME4Rec starts to increase at the beginning, and it gradually reaches its peak when $\alpha$ is 0.4 on Amazon-Beauty, 0.8 on Amazon-Clothing, and 0.3 on Amazon-Sports.
Afterward, it starts to decline.
Interestingly, the best $\alpha$ for the Sports dataset is only 0.3 and 0.4 for Beauty.
It suggests that for sparse datasets, the sliding filter size does not need to be too large. 
It is necessary to focus on the specific frequency bands where important frequency features exist.
For example in the Amazon dataset, the important frequency components of users are more concentrated and mainly distributed in the low-frequency region \cite{FMLP-Rec}.
While on dense datasets like ML-1M composed of various kinds of movies, the spectrum of user sequences is more complex and the important frequency components are scattered in various frequency bands, making users' interests in the ML-1M dataset more diverse and harder to understand.
To capture more complex sequential patterns of users, the dynamic filter structure requires more complex parameters in the frequency domain and a larger receptive field in the time domain.
Note that no matter the value of $\alpha$ except 0.1 and 1, our method SLIME4Rec always performs better than DuoRec.
However, when $\alpha$ is set too small ($\alpha$ = 0.1), it will lead to suboptimal recommendation results. 

\subsubsection{Impact of Input Sequence Length $N$}
\begin{figure*}[!t]
	\centering
	{\includegraphics[width=1\linewidth]{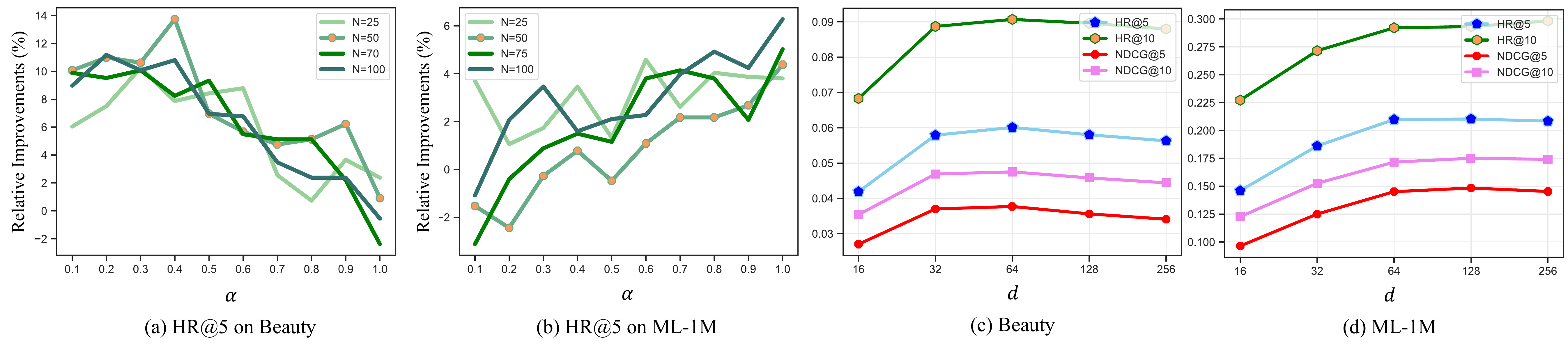}}
	\caption{Performances with different dynamic filter size ratio $\alpha$ when setting different max item list length. And performances with different hidden size $d$.}
	\label{fig:lengthandhidden}
\end{figure*}

As mentioned in section~\ref{method}, $\alpha$ is the ratio of the dynamic filter size $S_D$ relative to the frequency domain feature range $M$, where $M$ depends on $N$. 
To investigate the benefit of utilizing a sliding filter of appropriate size along the sequence length dimension, we conduct SLIME4Rec with different $N$. 
We only report the results on HR@5 due to space limitations, and similar trends can be found on other metrics.
Figure~\ref{fig:lengthandhidden}(a) and (b) show how the relative improvements of HR@5 vary with filter size ratio $\alpha$ when using different input sequence length.
For the Beauty dataset, increasing the $N$ of SLIME4Rec from 25 to 50 consistently improves the performance but when $N$ becomes larger, the model performance will decline.
These results suggest that increasing $N$ arbitrarily does not always result in an improvement in performance.
However, for the dense dataset ML-1M with more average actions of users, increasing the sequence length $N$ when the dynamic filter size ratio equals 1 is promised to improve performance since more existing interacted items are considered.
Finally, it can be seen that the dynamic filter size ratio is not sensitive to the change in sequence length despite increasing the sequence length.
\begin{table}[t]\Huge
\renewcommand{\arraystretch}{1.4}
    \centering
    \caption{
    Performance comparison of SLIME4Rec with DuoRec at different layers. HR is short for Hit Recal and NG is short for NDCG. The best performance in each row is boldfaced.
    }
    \label{tab:seq5}
    \begin{adjustbox}{max width=\linewidth}
            \begin{tabular}{cccccccccccccccc}
            \toprule
            \toprule
            \multicolumn{2}{c}{
            \multirow{2}{*}{Method}
            }
            &
            \multicolumn{4}{c}{\Huge{Beauty}} 
            &
            & 
            \multicolumn{4}{c}{Clothing}
            &
            & 
            \multicolumn{4}{c}{Sports}\\
            \cmidrule{3-6}
            \cmidrule{8-11}
            \cmidrule{13-16}
            \multicolumn{2}{c}{}
            & HR@5
            & NG@5
            & HR@10
            & NG@10
            &
            & HR@5
            & NG@5
            & HR@10
            & NG@10
            &
            & HR@5
            & NG@5
            & HR@10
            & NG@10\\
            
            \midrule
            \midrule
            \multirow{3}{*}{\textbf{$L=2$}} 
            & DuoRec
            & 0.0546
            & 0.0352
            & 0.0845
            & 0.0443
            &
            & 0.0193
            & 0.0113
            & 0.0302
            & 0.0148
            &
            & 0.0326
            & 0.0208
            & 0.0498
            & 0.0262\\

            & \multirow{2}{*}{\textbf{Ours}}
            & \textbf{0.0604}
            & \textbf{0.0370}
            & \textbf{0.0903}
            & \textbf{0.0467}
            &
            & \textbf{0.0225}
            & \textbf{0.0126}
            & \textbf{0.0343}
            & \textbf{0.0164}
            &
            & \textbf{0.0364}
            & \textbf{0.0230}
            & \textbf{0.0561}
            & \textbf{0.0294}\\

            &
            & $\uparrow$10.62\%
            & $\uparrow$5.11\%
            & $\uparrow$6.86\%
            & $\uparrow$5.42\%
            &
            & $\uparrow$16.58\% 
            & $\uparrow$11.50\%
            & $\uparrow$13.58\%
            & $\uparrow$10.81\%
            &
            & $\uparrow$13.75\%
            & $\uparrow$10.58\%
            & $\uparrow$13.79\%
            & $\uparrow$12.21\%\\
            
            \midrule
            \multirow{3}{*}{\textbf{$L=4$}} 
            & DuoRec
            & 0.0551
            & 0.0344
            & 0.0855
            & 0.0441
            &
            & 0.0197
            & 0.0113
            & 0.0299
            & 0.0146
            &
            & 0.0315
            & 0.0204
            & 0.0480
            & 0.0257
            \\

            & \multicolumn{1}{c}{\multirow{2}{*}{\textbf{Ours}}}
            & \textbf{0.0607}
            & \textbf{0.0379}
            & \textbf{0.0920}
            & \textbf{0.0480}
            &
            & \textbf{0.0221}
            & \textbf{0.0126}
            & \textbf{0.0341}
            & \textbf{0.0165}
            &
            & \textbf{0.0373}
            & \textbf{0.0243}
            & \textbf{0.0565}
            & \textbf{0.0305}
            \\

            & \multicolumn{1}{c}{}
            & $\uparrow$10.16\%
            & $\uparrow$10.17\%
            & $\uparrow$7.60\%
            & $\uparrow$8.84\%
            &
            & $\uparrow$12.18\%
            & $\uparrow$11.50\%
            & $\uparrow$14.05\%
            & $\uparrow$13.01\%
            &
            & $\uparrow$17.46\%
            & $\uparrow$19.12\%
            & $\uparrow$19.17\%
            & $\uparrow$18.68\%
            \\
            
            \midrule
            \multirow{3}{*}{\textbf{$L=8$}} 
            & DuoRec
            & 0.0565
            & 0.0353
            & 0.0867
            & 0.0451
            &
            & 0.0197
            & 0.0116
            & 0.0316
            & 0.0154
            &
            & 0.0299
            & 0.0197
            & 0.0460
            & 0.0248\\

            & \multicolumn{1}{c}{\multirow{2}{*}{\textbf{Ours}}}
            & \textbf{0.0621}
            & \textbf{0.0396}
            & \textbf{0.0910}
            & \textbf{0.0489}
            &
            & \textbf{0.0221}
            & \textbf{0.0128}
            & \textbf{0.0342}
            & \textbf{0.0167}
            &
            & \textbf{0.0365}
            & \textbf{0.0239}
            & \textbf{0.0563}
            & \textbf{0.0302}\\

            & \multicolumn{1}{c}{}
            & $\uparrow$9.91\%
            & $\uparrow$12.18\%
            & $\uparrow$4.96\%
            & $\uparrow$8.43\%
            &
            & $\uparrow$12.18\%
            & $\uparrow$10.34\%
            & $\uparrow$8.23\%
            & $\uparrow$8.44\%
            &
            & $\uparrow$22.07\%
            & $\uparrow$21.32\%
            & $\uparrow$19.17\%
            & $\uparrow$21.77\%\\
            
            \bottomrule
            \bottomrule
\end{tabular}   
    \end{adjustbox}
\end{table}
\begin{table}[t]\Huge
\renewcommand{\arraystretch}{1}
    \centering
    \label{tab:seq}
    \begin{adjustbox}{max width=\linewidth}
            \begin{tabular}{ccccccccccc}
            \toprule
            \toprule
            \multicolumn{2}{c}{\multirow{2}{*}{Method}}                               
            & \multicolumn{4}{c}{ML-1M} 
            &
            & \multicolumn{4}{c}{Yelp} \\
            \cmidrule{3-6}
            \cmidrule{8-11}
            \multicolumn{2}{c}{}
            & HR@5
            & NG@5
            & HR@10
            & NG@10
            &
            & HR@5
            & NG@5
            & HR@10
            & NG@10\\
            
            \midrule
            \midrule
            \multirow{3}{*}{\textbf{$L=2$}} 
            & DuoRec
            & 0.2038
            & 0.1390
            & 0.2946
            & 0.1680
            &
            & 0.0441
            & 0.0325
            & 0.0631
            & 0.0386\\

            & \multirow{2}{*}{\textbf{Ours}}
            & \textbf{0.2139}
            & \textbf{0.1457}
            & \textbf{0.3026}
            & \textbf{0.1743}
            &
            & \textbf{0.0516}
            & \textbf{0.0359}
            & \textbf{0.0766}
            & \textbf{0.0439}\\

            &
            & $\uparrow$4.96\%
            & $\uparrow$4.82\%
            & $\uparrow$2.72\%
            & $\uparrow$3.75\%
            &
            & $\uparrow$17.01\%
            & $\uparrow$10.46\%
            & $\uparrow$21.39\%
            & $\uparrow$13.7\%\\
            
            \midrule
            \multirow{3}{*}{\textbf{$L=4$}} 
            & DuoRec
            & 0.2065
            & 0.1423
            & 0.2917
            & 0.1699
            &
            & 0.0454
            & 0.0333
            & 0.0643
            & 0.0394\\

            & \multicolumn{1}{c}{\multirow{2}{*}{\textbf{Ours}}}
            & \textbf{0.2202} 
            & \textbf{0.1515}
            & \textbf{0.3127}
            & \textbf{0.1812}
            &
            & \textbf{0.0502} 
            & \textbf{0.0348}
            & \textbf{0.0765}
            & \textbf{0.0432}\\

            & \multicolumn{1}{c}{}
            & $\uparrow$6.63\%
            & $\uparrow$6.47\%
            & $\uparrow$7.20\%
            & $\uparrow$6.65\%
            &
            & $\uparrow$10.57\%
            & $\uparrow$4.5\%
            & $\uparrow$18.97\%
            & $\uparrow$9.64\%\\
            
            \midrule
            \multirow{3}{*}{\textbf{$L=8$}} 
            & DuoRec
            & 0.2164
            & 0.1501
            & 0.3063
            & 0.1701
            &
            & 0.0438
            & 0.0318
            & 0.0629
            & 0.0380\\

            & \multicolumn{1}{c}{\multirow{2}{*}{\textbf{Ours}}}
            & \textbf{0.2262} 
            & \textbf{0.1559}
            & \textbf{0.3132}
            & \textbf{0.1840}
            &
            & \textbf{0.0493} 
            & \textbf{0.0336}
            & \textbf{0.0745}
            & \textbf{0.0417}\\

            & \multicolumn{1}{c}{}
            & $\uparrow$4.53\%
            & $\uparrow$3.86\%
            & $\uparrow$2.25\%
            & $\uparrow$8.17\%
            &
            & $\uparrow$9.80\%
            & $\uparrow$5.66\%
            & $\uparrow$14.97\%
            & $\uparrow$9.74\%\\
            
            \bottomrule
            \bottomrule
\end{tabular}   
    \end{adjustbox}
\end{table}

\subsubsection{Impact of Hidden Size $d$}
We also conduct experiments on different hidden sizes $d$ in addition to the experiment of sequence length $N$ to fully examine the impact of filter size across all dimensions.
We vary the hidden size $d$ of our latent representations from 16 to 256.
The experiment results are displayed in Figure~\ref{fig:alpha}.
We observe that for user sequential behavior modeling, a limited dimension size cannot preserve enough latent information of items.
The model performance saturates as the number of hidden units reaches around 64 on Beauty and ML-1M since the larger embedding dimensionality enhances the representation of each frequency component with more dimensions.
But when the hidden size goes above the optimal point, the outcomes stop improving and even start to deteriorate.
This demonstrates that when the hidden size is too large, it may result in overfitting.

\subsubsection{Impact of Model Depth $L$}
The depth of the model determines the capacity of and the parameters of the backbone encoder, thus affecting the performance of the sequential representation learning.
Existing SR models are not very deep, however in our experiments we found that SLIME4Rec benefits from more layers.
To investigate the impact of filter mixer layer numbers, we fix other parameters and change the number of network layers for both SLIME4Rec and DuoRec to be the same for a fair comparison.
Table~\ref{tab:seq5} demonstrates the experimental result.
We can observe that SLIME4Rec achieves the best performance with the setting $L=8$ on Beauty and ML-1M dataset while getting the best result with the setting $L=2$ or $L=4$ on the Clothing, Sports, and Yelp dataset.
In general, SLIME4Rec constantly outperforms baseline under all different network depth settings and can stack more layers without losing too much recommendation performance, which is largely due to using smaller filters and allowing each layer to focus on capturing user behavior patterns in a specific frequency range.
This shows that the boost in the performance of the Transformer cannot be achieved by simply adding more layers, while the proposed slide filter mixer structure utilizes dynamic and static components wisely and can capture higher interaction order of the features while avoiding overfitting.


\begin{figure}[!t]
	\centering
	{\includegraphics[width=1\linewidth]{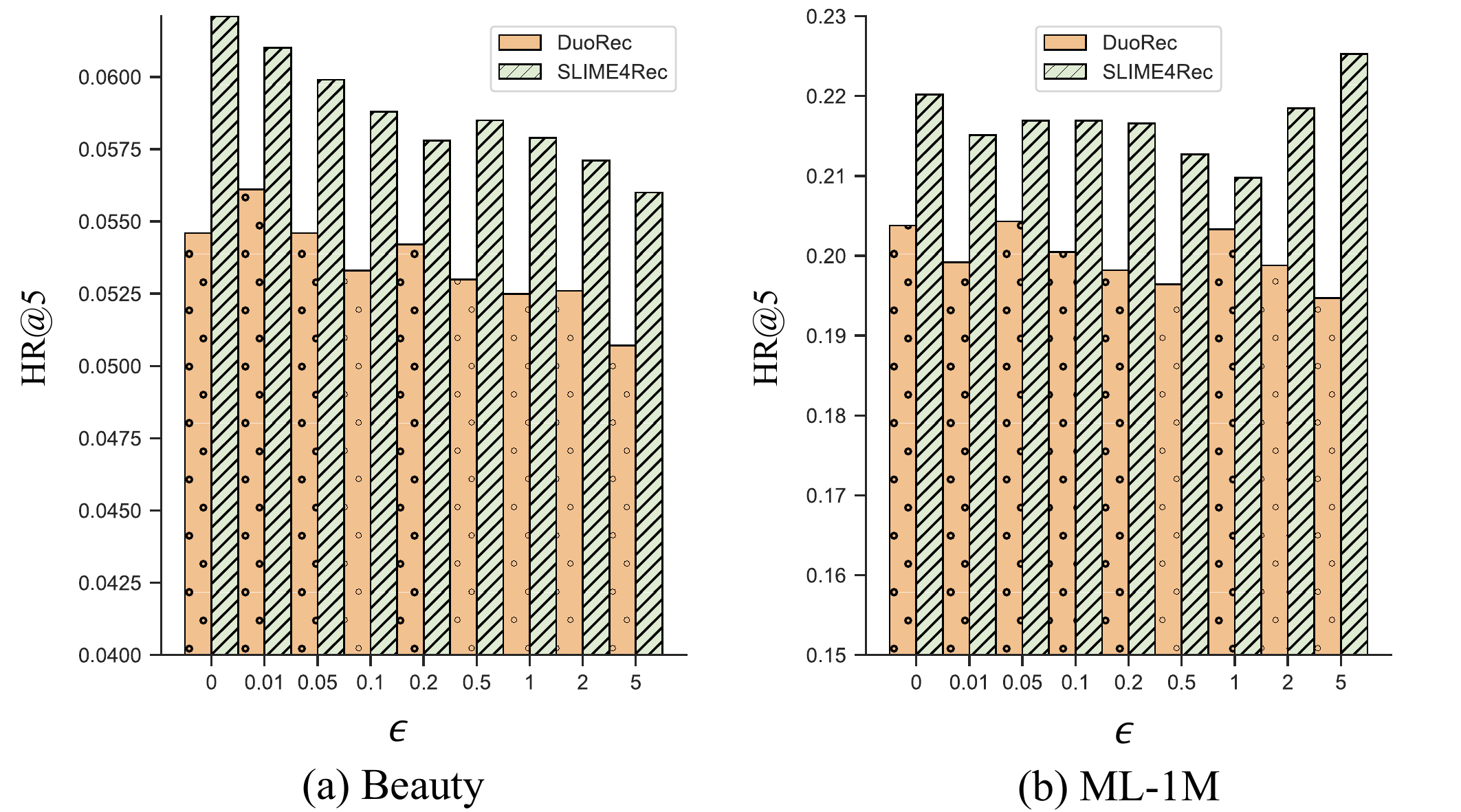}}
	\caption{Robustness to different proportion of synthetic noise.}
	\label{fig:noise}
\end{figure}
\subsection{Robustness to Synthetic Noises}
To verify the robustness of SLIME4Rec to synthetic noises (\textbf{RQ5}), we conduct experiments on the Beauty and ML-1M datasets.
Specifically, random uniform noises are added to the original representations at each layer.
From Figure~\ref{fig:noise}, we can see that the increase of $\epsilon$ leads to the degradation of the performance of SLIME4Rec and DuoRec.
However, the performance of SLIME4Rec is consistently higher than that of DuoRec.
The reason might be the noises added in the time domain can be easily distinguished by our slide filter in the frequency domain.
Interestingly, on the dense ML-1M dataset, when the proportion of added noise is too large, the recommendation performance of SLIME4Rec does not decrease but improves a lot. 
The reason might be that excessive noise can be easily removed from the spectrum of dense datasets, and better representations without noise are generated for contrastive learning.
In other words, SLIME4Rec shows great robustness to random noise.
\begin{figure}[!t]
	\centering
	{\includegraphics[width=1\linewidth]{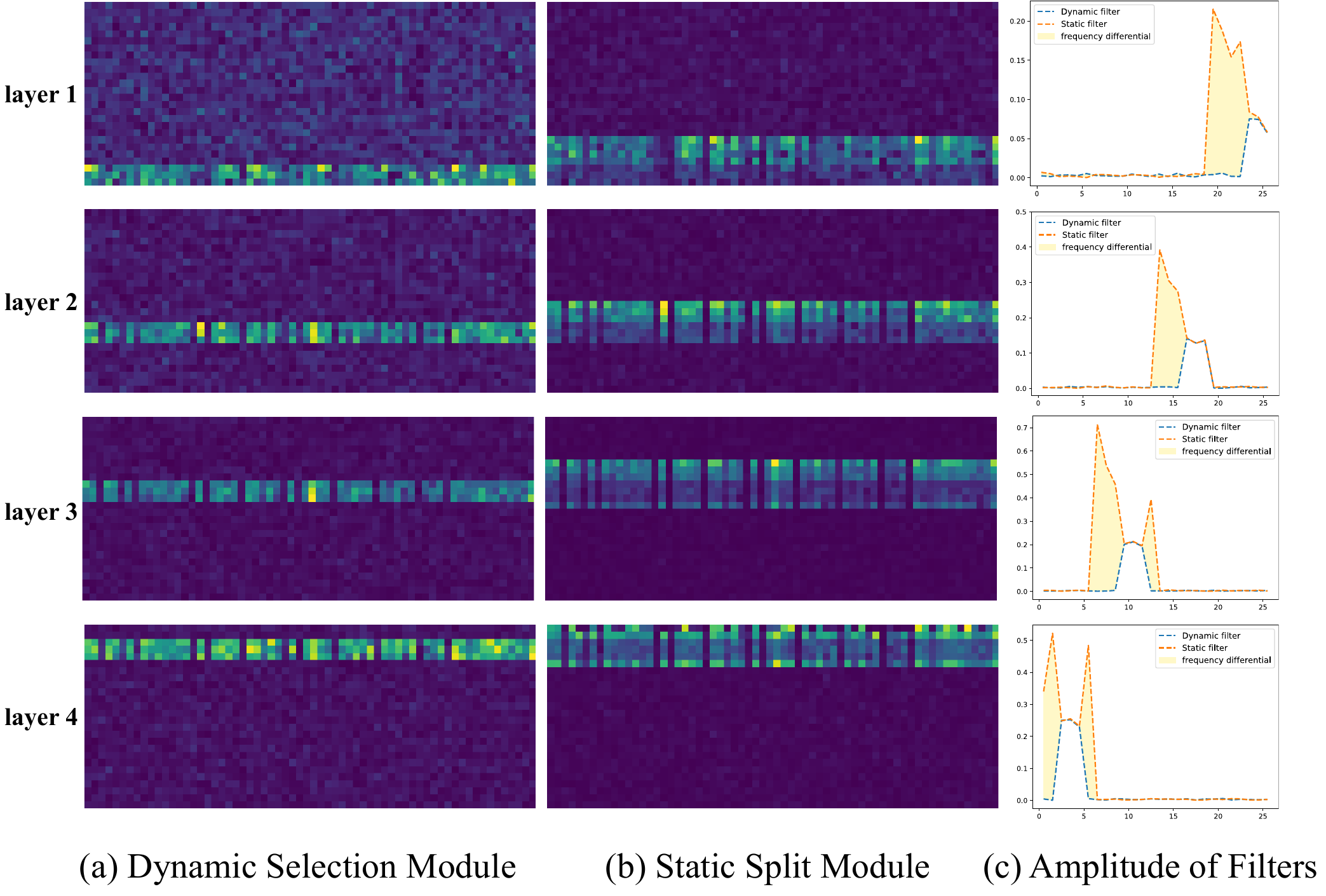}}
	\caption{Visualization of the learned slide filters in SLIME4Rec. We visualize the dynamic filters in (a) and show the corresponding static filters in (b) setting the slide mode four ($\leftarrow$) of frequency ramp structure with $\alpha=0.1$ and $\beta=0.25$.
	We compute the amplitude of dynamic and static filters along the sequence dimension and show the frequency differential between them in (c). There are some frequency patterns that are missed when $\alpha < \frac{1}{L}$ while using the static filter module can recapture them.}
	\label{fig:filter}
\end{figure}
\subsection{Visualization of Learned Filters}
To evaluate how the slide filter capture sequential behavior in each layer, the complex weights learned in the sliding filter are visualized on the Beauty dataset.
The visualization of dynamic and static filters can help understand how the frequency ramp structure works.
The results are displayed in Figure~\ref{fig:filter}.
Based on our observations from Figure~\ref{fig:filter}(a) and (b), we can derive the following results: 
All the dynamic and static filters of different layers have a clear frequency pattern which means that different layers focus on capturing the user’s patterns in different frequency ranges.
When capturing the characteristics of the current specific frequency range, the learnable filter will simply disable surrounding frequency components.
If we look at the whole picture, it is obvious to find that filters are sliding in mode 4, the dynamic and static filter slides from high-frequency to low-frequency features.

As mentioned in Section~\ref{method}, when the dynamic filter size $S_D < step$, 
there are some missing frequency components between adjacent frequency bands, which prevents our model from modeling the user's multi-term preference.
The frequency differential is illustrated in Figure~\ref{fig:filter}(c).
The static filter recaptures the frequency patterns missed by the dynamic filter since the static module splits the frequency components averagely by model depth $L$.
It also provides a more fine-grained frequency division, which helps to model users' complex preferences when these features are mixed with those extracted by the dynamic module.
Although the filter tends to assign more weight to the low frequency components, from the perspective of the whole frequency range, these relatively high frequency components are emphasized, which alleviates the problem that high frequency information is often overlooked.

\section{RELATED WORK}
\subsection{Sequential Recommendation}
Sequential recommendation forecasts future items throughout the user sequence by modeling item transition correlations.
Early SR studies are often based on the Markov Chain assumption.
Afterward, numerous sequential recommender models powered by deep learning were created.
GRU4Rec\cite{GRU4Rec} is the very first attempt to utilize the GRU network in SR. 
To discover sequential patterns, Caser\cite{Caser} employs both horizontal and vertical convolutional filters.
Later on, self-attention networks have shown great potential in modeling sequence data and a variety of related models are developed, e.g., SASRec \cite{SASRec}, BERT4Rec \cite{BERT4Rec} and S${^3}$Rec \cite{S3Rec}.
Most of these methods \cite{related28,related15,related7,related5,related18,related3, ouricde} use the next-item supervised training style as their training scheme.
The other training scheme usually has extra auxiliary training tasks.
CL4SRec \cite{CL4SRec} applies a contrastive strategy to multiple views generated by data augmentation.
CoSeRec \cite{CoSeRec} introduces more augmentation operations to train robust sequence representations. 
DuoRec \cite{DuoRec} combines recommendation loss with unsupervised learning and supervised contrastive learning to optimize the SR models.
Despite the success of these model in SR, they ignored important information hidden in the frequency domain.
\subsection{Frequency Domain Learning}
Fourier transformation has been an important tool in digital signal processing for decades~\cite{related35, related1}.
There are a variety of works that incorporate Fourier transformation in computer vision \cite{learninginFD, fastfourierconvolution, resolution, FFC-SE} and natural language processing \cite{NLP1, FNET}.
Global filter networks (GFNet) \cite{GFNet} learn Fourier filters to perform depthwise global convolution.
Very recent works try to leverage Fourier transform enhanced model for long-term series forecasting \cite{autoformer, FEDformer} and partial differential equations solving \cite{FNO, AFNO, UFNO}. 
However, there are few Fourier-related works in the sequential recommendation.
More recently, FMLP-Rec \cite{FMLP-Rec} utilizes an all-MLP structure without self-attention mechanism for SR.
More recently, FMLP-Rec \cite{FMLP-Rec} first introduce a filter-enchanced MLP for SR, which multiplies a global filter to remove noise in the frequency domain.
However, the global filter tend to give greater weight to low frequencies and underrate relatively high frequencies.

\section{CONCLUSION}
In this paper, we proposed a novel model SLIME4Rec for the sequential recommendation, which goes beyond the existing time feature-based methods in SR.
We built two frequency selection modules with frequency ramp structure namely dynamic frequency selection and static frequency split module, respectively, 
to find more fine-grained sequential patterns in different frequencies bands, 
thus alleviating the problem caused by noise. 
Finally, to the best of our knowledge, we are the first to employ a contrastive learning paradigm for frequency-based model to supplement primary recommendation loss with contrastive loss. 
Experimental results on five benchmark datasets showed the superiority of our SLIME4Rec model over all state-of-the-art models.
\section*{Acknowledgement}
This research was partially supported by the NSFC (61876117, 62176175), the major project of natural science research in Universities of Jiangsu Province (21KJA520004), Suzhou Science and Technology Development Program (SYC2022139), the Priority Academic Program Development of Jiangsu Higher Education Institutions and the Exploratory Self-selected Project of the State Key Laboratory of Software Development Environment.
\bibliographystyle{IEEEtran}
\bibliography{references}{}

\begin{thebibliography}{10}
\providecommand{\url}[1]{#1}
\csname url@samestyle\endcsname
\providecommand{\newblock}{\relax}
\providecommand{\bibinfo}[2]{#2}
\providecommand{\BIBentrySTDinterwordspacing}{\spaceskip=0pt\relax}
\providecommand{\BIBentryALTinterwordstretchfactor}{4}
\providecommand{\BIBentryALTinterwordspacing}{\spaceskip=\fontdimen2\font plus
\BIBentryALTinterwordstretchfactor\fontdimen3\font minus
  \fontdimen4\font\relax}
\providecommand{\BIBforeignlanguage}[2]{{%
\expandafter\ifx\csname l@#1\endcsname\relax
\typeout{** WARNING: IEEEtran.bst: No hyphenation pattern has been}%
\typeout{** loaded for the language `#1'. Using the pattern for}%
\typeout{** the default language instead.}%
\else
\language=\csname l@#1\endcsname
\fi
#2}}
\providecommand{\BIBdecl}{\relax}
\BIBdecl

\bibitem{BPR_MF}
S.~Rendle, C.~Freudenthaler, Z.~Gantner, and L.~Schmidt-Thieme, ``Bpr: Bayesian
  personalized ranking from implicit feedback,'' \emph{arXiv preprint
  arXiv:1205.2618}, 2012.

\bibitem{GRU4Rec}
D.~Jannach and M.~Ludewig, ``When recurrent neural networks meet the
  neighborhood for session-based recommendation,'' in \emph{RecSys}, 2017, pp.
  306--310.

\bibitem{Caser}
J.~Tang and K.~Wang, ``Personalized top-n sequential recommendation via
  convolutional sequence embedding,'' in \emph{WSDM}, 2018, pp. 565--573.

\bibitem{SASRec}
W.-C. Kang and J.~McAuley, ``Self-attentive sequential recommendation,'' in
  \emph{ICDM}.\hskip 1em plus 0.5em minus 0.4em\relax IEEE, 2018, pp. 197--206.

\bibitem{FMLP-Rec}
K.~Zhou, H.~Yu, W.~X. Zhao, and J.-R. Wen, ``Filter-enhanced mlp is all you
  need for sequential recommendation,'' in \emph{WWW}, 2022, pp. 2388--2399.

\bibitem{GFNet}
Y.~Rao, W.~Zhao, Z.~Zhu, J.~Lu, and J.~Zhou, ``Global filter networks for image
  classification,'' in \emph{NeurIPS 2021}, 2021, pp. 980--993.

\bibitem{DFT1}
L.~R. Rabiner and B.~Gold, ``Theory and application of digital signal
  processing,'' \emph{Englewood Cliffs: Prentice-Hall}, 1975.

\bibitem{DFT2}
S.~S. Soliman and M.~D. Srinath, ``Continuous and discrete signals and
  systems,'' \emph{Englewood Cliffs}, 1990.

\bibitem{howVITwork}
N.~Park and S.~Kim, ``How do vision transformers work?'' \emph{arXiv preprint
  arXiv:2202.06709}, 2022.

\bibitem{VIT_seelike_CNN}
M.~Raghu, T.~Unterthiner, S.~Kornblith, C.~Zhang, and A.~Dosovitskiy, ``Do
  vision transformers see like convolutional neural networks?'' \emph{NeurIPS},
  vol.~34, pp. 12\,116--12\,128, 2021.

\bibitem{FEDformer}
T.~Zhou, Z.~Ma, Q.~Wen, X.~Wang, L.~Sun, and R.~Jin, ``Fedformer: Frequency
  enhanced decomposed transformer for long-term series forecasting,'' in
  \emph{ICML}, ser. Proceedings of Machine Learning Research, vol. 162, 2022,
  pp. 27\,268--27\,286.

\bibitem{cooley}
J.~W. Cooley and J.~W. Tukey, ``An algorithm for the machine calculation of
  complex fourier series,'' \emph{Mathematics of computation}, vol.~19, no.~90,
  pp. 297--301, 1965.

\bibitem{FFTW3}
M.~Frigo and S.~G. Johnson, ``The design and implementation of fftw3,''
  \emph{Proceedings of the IEEE}, vol.~93, no.~2, pp. 216--231, 2005.

\bibitem{Dropout}
N.~Srivastava, G.~Hinton, A.~Krizhevsky, I.~Sutskever, and R.~Salakhutdinov,
  ``Dropout: a simple way to prevent neural networks from overfitting,''
  \emph{The journal of machine learning research}, vol.~15, no.~1, pp.
  1929--1958, 2014.

\bibitem{iFormer}
C.~Si, W.~Yu, P.~Zhou, Y.~Zhou, X.~Wang, and S.~Yan, ``Inception transformer,''
  \emph{arXiv preprint arXiv:2205.12956}, 2022.

\bibitem{CL4SRec}
X.~Xie, F.~Sun, Z.~Liu, S.~Wu, J.~Gao, J.~Zhang, B.~Ding, and B.~Cui,
  ``Contrastive learning for sequential recommendation,'' in \emph{ICDE}.\hskip
  1em plus 0.5em minus 0.4em\relax IEEE, 2022, pp. 1259--1273.

\bibitem{DuoRec}
R.~Qiu, Z.~Huang, H.~Yin, and Z.~Wang, ``Contrastive learning for
  representation degeneration problem in sequential recommendation,'' in
  \emph{WSDM}, 2022, pp. 813--823.

\bibitem{conjugate}
J.~W. Cooley and J.~W. Tukey, ``An algorithm for the machine calculation of
  complex fourier series,'' \emph{Mathematics of computation}, vol.~19, no.~90,
  pp. 297--301, 1965.

\bibitem{Amazon}
J.~McAuley, C.~Targett, Q.~Shi, and A.~Van Den~Hengel, ``Image-based
  recommendations on styles and substitutes,'' in \emph{SIGIR}, 2015, pp.
  43--52.

\bibitem{ML-1M}
F.~M. Harper and J.~A. Konstan, ``The movielens datasets: History and
  context,'' \emph{Acm transactions on interactive intelligent systems (tiis)},
  vol.~5, no.~4, pp. 1--19, 2015.

\bibitem{Timeinterval}
J.~Li, Y.~Wang, and J.~McAuley, ``Time interval aware self-attention for
  sequential recommendation,'' in \emph{WSDM}, 2020, pp. 322--330.

\bibitem{BERT4Rec}
F.~Sun, J.~Liu, J.~Wu, C.~Pei, X.~Lin, W.~Ou, and P.~Jiang, ``Bert4rec:
  Sequential recommendation with bidirectional encoder representations from
  transformer,'' in \emph{CIKM}, 2019, pp. 1441--1450.

\bibitem{kdd}
W.~Krichene and S.~Rendle, ``On sampled metrics for item recommendation,'' in
  \emph{KDD}, 2020.

\bibitem{ContrastVAE}
Y.~Wang, H.~Zhang, Z.~Liu, L.~Yang, and P.~S. Yu, ``Contrastvae: Contrastive
  variational autoencoder for sequential recommendation,'' in \emph{CIKM},
  2022, pp. 2056--2066.

\bibitem{CoSeRec}
Z.~Liu, Y.~Chen, J.~Li, P.~S. Yu, J.~McAuley, and C.~Xiong, ``Contrastive
  self-supervised sequential recommendation with robust augmentation,''
  \emph{arXiv preprint arXiv:2108.06479}, 2021.

\bibitem{S3Rec}
K.~Zhou, H.~Wang, W.~X. Zhao, Y.~Zhu, S.~Wang, F.~Zhang, Z.~Wang, and J.-R.
  Wen, ``S3-rec: Self-supervised learning for sequential recommendation with
  mutual information maximization,'' in \emph{CIKM}, 2020, pp. 1893--1902.

\bibitem{related28}
L.~Wu, S.~Li, C.-J. Hsieh, and J.~Sharpnack, ``Sse-pt: Sequential
  recommendation via personalized transformer,'' in \emph{RecSys}, 2020, pp.
  328--337.

\bibitem{related15}
J.~Lin, W.~Pan, and Z.~Ming, ``Fissa: fusing item similarity models with
  self-attention networks for sequential recommendation,'' in \emph{RecSys},
  2020, pp. 130--139.

\bibitem{related7}
Y.~He, Y.~Zhang, W.~Liu, and J.~Caverlee, ``Consistency-aware recommendation
  for user-generated item list continuation,'' in \emph{WSDM}, 2020, pp.
  250--258.

\bibitem{related5}
X.~Fan, Z.~Liu, J.~Lian, W.~X. Zhao, X.~Xie, and J.-R. Wen, ``Lighter and
  better: low-rank decomposed self-attention networks for next-item
  recommendation,'' in \emph{SIGIR}, 2021, pp. 1733--1737.

\bibitem{related18}
Z.~Liu, Z.~Fan, Y.~Wang, and P.~S. Yu, ``Augmenting sequential recommendation
  with pseudo-prior items via reversely pre-training transformer,'' in
  \emph{SIGIR}, 2021, pp. 1608--1612.

\bibitem{related3}
Z.~Cui, Y.~Cai, S.~Wu, X.~Ma, and L.~Wang, ``Motif-aware sequential
  recommendation,'' in \emph{SIGIR}, 2021, pp. 1738--1742.

\bibitem{ouricde}
J.~Zhao, P.~Zhao, L.~Zhao, Y.~Liu, V.~S. Sheng, and X.~Zhou, ``Variational
  self-attention network for sequential recommendation,'' in \emph{ICDE}.\hskip
  1em plus 0.5em minus 0.4em\relax IEEE, 2021, pp. 1559--1570.

\bibitem{related35}
I.~Pitas, \emph{Digital image processing algorithms and applications}.\hskip
  1em plus 0.5em minus 0.4em\relax John Wiley \& Sons, 2000.

\bibitem{related1}
G.~A. Baxes, \emph{Digital image processing: principles and
  applications}.\hskip 1em plus 0.5em minus 0.4em\relax John Wiley \& Sons,
  Inc., 1994.

\bibitem{learninginFD}
K.~Xu, M.~Qin, F.~Sun, Y.~Wang, Y.-K. Chen, and F.~Ren, ``Learning in the
  frequency domain,'' in \emph{CVPR}, 2020, pp. 1740--1749.

\bibitem{fastfourierconvolution}
L.~Chi, B.~Jiang, and Y.~Mu, ``Fast fourier convolution,'' \emph{NeurIPS},
  vol.~33, pp. 4479--4488, 2020.

\bibitem{resolution}
R.~Suvorov, E.~Logacheva, A.~Mashikhin, A.~Remizova, A.~Ashukha, A.~Silvestrov,
  N.~Kong, H.~Goka, K.~Park, and V.~Lempitsky, ``Resolution-robust large mask
  inpainting with fourier convolutions,'' in \emph{WACV}, 2022, pp. 2149--2159.

\bibitem{FFC-SE}
I.~Shchekotov, P.~Andreev, O.~Ivanov, A.~Alanov, and D.~Vetrov, ``Ffc-se: Fast
  fourier convolution for speech enhancement,'' \emph{arXiv preprint
  arXiv:2204.03042}, 2022.

\bibitem{NLP1}
A.~Tamkin, D.~Jurafsky, and N.~Goodman, ``Language through a prism: A spectral
  approach for multiscale language representations,'' \emph{NeurIPS}, vol.~33,
  pp. 5492--5504, 2020.

\bibitem{FNET}
J.~Lee-Thorp, J.~Ainslie, I.~Eckstein, and S.~Ontanon, ``Fnet: Mixing tokens
  with fourier transforms,'' \emph{arXiv preprint arXiv:2105.03824}, 2021.

\bibitem{autoformer}
H.~Wu, J.~Xu, J.~Wang, and M.~Long, ``Autoformer: Decomposition transformers
  with auto-correlation for long-term series forecasting,'' \emph{NeurIPS},
  vol.~34, pp. 22\,419--22\,430, 2021.

\bibitem{FNO}
Z.~Li, N.~Kovachki, K.~Azizzadenesheli, B.~Liu, K.~Bhattacharya, A.~Stuart, and
  A.~Anandkumar, ``Fourier neural operator for parametric partial differential
  equations,'' \emph{arXiv preprint arXiv:2010.08895}, 2020.

\bibitem{AFNO}
J.~Guibas, M.~Mardani, Z.~Li, A.~Tao, A.~Anandkumar, and B.~Catanzaro,
  ``Adaptive fourier neural operators: Efficient token mixers for
  transformers,'' \emph{arXiv preprint arXiv:2111.13587}, 2021.

\bibitem{UFNO}
G.~Wen, Z.~Li, K.~Azizzadenesheli, A.~Anandkumar, and S.~M. Benson,
  ``U-fno—an enhanced fourier neural operator-based deep-learning model for
  multiphase flow,'' \emph{Advances in Water Resources}, vol. 163, p. 104180,
  2022.

\end{thebibliography}
\end{document}